\UseRawInputEncoding 
\documentclass[5p]{elsarticle}

\usepackage[version=3]{mhchem}
\usepackage{siunitx}
\usepackage{graphicx}
\usepackage{amsmath}
\usepackage{color, colortbl} \usepackage{subfig} \usepackage{float}
\usepackage{booktabs}

\usepackage{algorithm} \usepackage{algpseudocode} \usepackage{pifont}

\usepackage[draft]{changes} \definechangesauthor[name={Valerio},color={red}]{v}
\usepackage{todonotes}

\newcolumntype{P}{>{\raggedleft\arraybackslash}p{.2in}}

\definecolor{hdbg}{RGB}{57,81,117} \definecolor{LRed}{rgb}{1,.8,.8}
\definecolor{Gray}{gray}{0.9}

\begin{document}
\sloppy

\title{A Personalized Recommender System for Pervasive Social Networks}

\author[iit]{Valerio Arnaboldi} \ead{v.arnaboldi@iit.cnr.it} \author[iit]{Mattia
  G. Campana\corref{cor1}} \ead{m.campana@iit.cnr.it} \author[iit]{Franca
  Delmastro} \ead{f.delmastro@iit.cnr.it} \author[mil,iit]{Elena Pagani}
\ead{pagani@di.unimi.it} \cortext[cor1]{Corresponding author}
\address[iit]{IIT-CNR, Via G. Moruzzi 1, 56124, Pisa, Italy}
\address[mil]{Department of Computer Science, Universit\`a degli Studi di
  Milano, Via Comelico 39, 20135 Milano, Italy}

\begin{abstract}

The current availability of interconnected portable devices, and the advent of
the Web 2.0, raise the problem of supporting anywhere and anytime access to a
huge amount of content, generated and shared by mobile users. On the one hand,
users tend to be always connected for sharing experiences and conducting their
social interactions with friends and acquaintances, through so-called Mobile
Social Networks, further improving their social inclusion. On the other hand,
the pervasiveness of communication infrastructures spreading data (cellular
networks, direct device-to-device contacts, interactions with ambient devices as
in the Internet-of-Things) makes compulsory the deployment of solutions able to
filter off undesired information and to select what content should be addressed
to which users, for both $(i)$ better user experience, and $(ii)$ resource
saving of both devices and network.

In this work, we propose a novel framework for pervasive social networks, called
{\it Pervasive PLIERS (p-PLIERS)}, able to discover and select, in a highly
personalized way, contents of interest for single mobile users. p-PLIERS
exploits the recently proposed PLIERS tag-based recommender
system~\cite{arnaboldi2016pliers} as context a reasoning tool able to adapt
recommendations to heterogeneous interest profiles of different users. p-PLIERS
effectively operates also when limited knowledge about the network is
maintained. It is implemented in a completely decentralized environment, in
which new contents are continuously generated and diffused through the network,
and it relies only on the exchange of single nodes' knowledge during proximity
contacts and through device-to-device communications. We evaluated p-PLIERS by
simulating its behavior in three different scenarios: a big event (Expo 2015), a
conference venue (ACM KDD'15), and a working day in the city of Helsinki. For
each scenario, we used real or synthetic mobility traces and we extracted real
datasets from Twitter interactions to characterize the generation and sharing of
user contents.
 
\end{abstract}


\maketitle

\section{Introduction}

The amount of data accessible through the Internet has dramatically increased
over the last years. This trend has been exacerbated by the advent of online
social networks (hereinafter OSNs) and by video-on-demand services like
Netflix~\cite{bennett2007netflix}. It is then expected to boom with the Internet
of Things, where potentially every object in the physical world will create and
share information over the network.

The availability of this data represents an important resource that is
revolutionizing our society, but it is also posing some serious technological
challenges in terms of maintenance, management, indexing and identification of
contents. This affects especially mobile communications, where users want to be
always connected and able to share contents anywhere and anytime. To alleviate
the burden of data traffic on cellular networks, several solutions based on
device-to-device (D2D) wireless communications have been proposed in the
literature, such as opportunistic networks~\cite{Conti2010}. However, most of
the classical mechanisms for content identification and recommendation designed
for centralized infrastructures cannot be directly applied to these scenarios,
and ad-hoc solutions must be adopted. This is the case of Mobile Social Networks
(hereinafter MSNs), designed to further improve social interactions and
inclusion through experience sharing based on opportunistic communications, and
to this aim they need efficient and personalized mechanisms for useful content
selection and distribution.

The basic approaches to identify useful contents in opportunistic networks are
based on publish/subscribe mechanisms. Users have to explicitly define their
interests by subscribing to a fixed set of thematic channels and, when they
encounter other mobile users, they can ask them for contents related to the
channels they are subscribed to (see for example the PodNet
project~\cite{lend07}). Other solutions exploit context information (e.g., the
history of physical contacts between nodes, social information about the users
or the presence of communities) to improve forwarding and content dissemination
to potentially interested users (ContentPlace~\cite{boldrini2008csa},
ProfileCast~\cite{hsu2008profile}, SocialCast~\cite{costa2008socially} and
ICast~\cite{pagani2015weak}). They are generally defined as context-aware
forwarding or content dissemination protocols, and context information generally
includes information characterizing the user's behavior, her interests,
generated and shared contents and the surrounding environment.

Recently researchers have investigated the possibility of using recommender
systems~\cite{kantor2011recommender} to identify useful contents also in the
opportunistic environment (mainly based on content filtering and tag
expansion)~\cite{del2010differs,schifanella2008mobhinter,seshadri2013mobile,lo2010folksonomy}. Recommender
systems perform better than publish/subscribe mechanisms by mainly relying on
information about users' past actions (e.g., past purchases in
e-commerce~\cite{kohavi2001applications, linden2003amazon} or past
visualizations of multimedia content in video-on-demand
services~\cite{bennett2007netflix, ali2004tivo}). At the same time, they can be
included in the more general definition of context-aware systems, in which
context information is mainly focused on content characterization and it is
directly defined by users.  In fact, through the use of Social Tagging Systems
(STS), mobile users can directly define {\it tags} (i.e., labels) that describe
contents from a semantic point of view. The ensemble of tags generated by users
in a certain online system is known in the literature as
\textit{folksonomy}~\cite{lo2010folksonomy, mathes2004folksonomies}. An
important aspect of folksonomies is that, differently from ontologies, no
relationship between the terms is required a priori (hierarchical or not). On
the contrary, these relationships are automatically built thanks to the tags
created by the users and assigned to contents. Folksonomies have the ability to
quickly adapt to changes in the user's vocabulary and to represent highly
personalized information about the users' interests. These aspects fit well
mobile users' behavior, especially when they define, generate, and share
contents on-the-fly, while participating in an event or living a particular
experience.

The richness of information contained in folksonomies can be used to improve the
identification of interesting contents for each single user, and also to
determine relationships and affinity between different users, depending on the
content they generate and share. To this aim, it is essential to define an
algorithm able to efficiently detect interesting content starting from the local
user preferences and a limited knowledge of what is available on the
network. This represents a context reasoning problem in a distributed and mobile
environment, where each mobile device has a \emph{local} representation of
context information in the network (e.g., folksonomy in case of recommender
systems). This local knowledge is a view of the \emph{global} knowledge of the
information available from the whole network. Nodes can enrich their local
knowledge by sharing it with other nodes they physically encounter, through
opportunistic communications. They can take decisions about which contents to
download or forward to other nodes using their local knowledge. In this way,
mobile users can identify interesting content locally, without accessing a
centralized service.

However, since folksonomies are based on user-defined tags, they also have some
drawbacks. Synonyms, homonyms, polysemies, and different users' tagging behavior
make the reasoning process difficult to perform in some cases, and undermine the
use of simple tag matching~\cite{heymann2008can}. To overcome these limitations,
a novel family of recommender systems, called \textit{Tag-based Recommender
  Systems}~\cite{zhang2011tag}, have been recently proposed for centralized
infrastructures, where global knowledge is available. To understand whether a
content is interesting for a user, tag-based recommender systems do not only
consider tags that are directly associated with contents, but also the relations
existing between tags, trying to extract a semantic meaning from the
contents. To the best of our knowledge, currently there is no solution in the
literature proposing the use of tag-based recommender systems in opportunistic
networks. We think that this could substantially change the way mobile users and
services can access contents, especially in Mobile Social Network scenarios.

We recently proposed a novel tag-based recommender system, called
PLIERS~\cite{arnaboldi2016pliers}, which outperforms the state-of-the-art
tag-based recommender systems defined for online social networks and centralized
solutions. In this paper, we present a novel framework that exploits PLIERS
principles in a completely decentralized environment, relying only on the
exchange of single nodes' knowledge during proximity contacts and through D2D
communications. We called it {\it Pervasive PLIERS (p-PLIERS)}. It represents a
general framework for identifying useful and interesting contents in the mobile
environment, on top of which several services can be built~--~from context-aware
forwarding protocols, to content dissemination services (e.g., targeted
advertising), to content sharing services, etc. We extensively evaluated the
proposed solution through simulations considering three different application
scenarios: (i) users attending Expo2015 during the World Food Day (one of the
most crowded day of the entire event); (ii) users attending a scientific
conference (ACM KDD 2015); (iii) users moving around the city of Helsinki during
a working day. For each scenario, we selected appropriate mobility traces,
synthetic or real (when available) and we extracted real datasets from
geo-localized Twitter interactions. The datasets would reflect the behavior of
mobile users, generating and sharing contents related to a specific event or,
more in general, to their life in a European city.

The rest of the paper is organized as follows. In Section~\ref{sec:relwork}, we
present a summary of the use of recommender systems in opportunistic networks
for the identification of interesting contents. In Section~\ref{pliers}, we
describe PLIERS and we compare it with existing tag-based recommender
systems. Then, in Section~\ref{the_algorithm}, we present p-PLIERS framework in
detail. In Section~\ref{sec:experiments_gen}, we provide a general description
of the experimental scenarios that we considered for the evaluation of the
proposed solution. Specifically, in Section~\ref{sec:experiments_static}, we
present a comparison among existing recommender systems used in pervasive social
networks, showing the advantages of PLIERS. In
Section~\ref{sec:experiments_dynamic}, we present p-PLIERS performance
evaluation through simulations in the three different scenarios. Finally, we
discuss examples of services that can benefit from the framework in
Section~\ref{sec:possible_applications}, and we conclude the paper in
Section~\ref{sec:conclusion}.

\section{Related Work}
\label{sec:relwork}

In the last few years, many researchers have used Recommender Systems
for disseminating content in a mobile and pervasive setting. Most of the
approaches in the literature are based on popular Web Recommender Systems,
conveniently modified for mobile environments~--~e.g., by reducing the
computational complexity of the algorithms and limiting the amount of necessary
memory.

The most popular and widely implemented system is represented by the
\textit{Collaborative Filtering} (CF)~\cite{herlocker2004evaluating}
approach. The simplest implementation of CF makes recommendations to a user
based on items that other users with similar interests liked in the past. The
similarity in interests between two users is calculated by a similarity metric
(e.g., cosine similarity or Pearson correlation) between their respective
histories (i.e., the sets of items they liked in the past). This kind of CF is
also known as ``\textit{user-based CF}'', in contrast to the
``\textit{item-based CF}'' which models the preference of a user for an item
based on ratings of similar items by the same user.

Typically, CF-based systems operate on second-order tensors (or matrices) that
represent the relationships between users and items (or an item-item matrix for
the item-based CF). For this reason, CF systems could suffer from scalability
problems depending on the size of the data structures to be kept in memory and
on the sparseness degree of the matrix. Since the number of items in STS is
typically high and far beyond users' ability to evaluate even a small fraction
of them, the data representation in a CF matrix is often highly sparse. Recent
research work (e.g.,
\cite{del2010differs,schifanella2008mobhinter,seshadri2013mobile}) focused on
the reduction of the complexity of CF in order to identify useful content for
mobile users in opportunistic networks and then optimize content
dissemination. For example, \textit{diffeRS}~\cite{del2010differs} tries to
classify users in two separate classes: \textit{mass-like minded} and
\textit{atypical} users. The former set is composed by users whose preferences
are similar to the interests of other users encountered in the past (i.e., the
community). For this kind of users, the recommendation is simply based on the
average of the community's preferences. For atypical users (i.e., users whose
preferences differ from those of the community), instead, a user-based CF
approach is applied, computed only among those other atypical users who
similarly deviate from the community. Thereby, nodes exchange only the contents
that CF identifies as attractive to the local
users. MobHinter~\cite{schifanella2008mobhinter}, instead, limits the CF
computation to the most similar users according to a certain similarity measure,
which takes into account different parameters (e.g., resources in common,
similar behaviors, and so on). Moreover, it proposes different strategies to
limit the amount of information exchanged by nodes every time they meet.
However, even though there has been a complexity reduction to allow CF to run on
mobile devices with limited computational resources, these approaches take into
account just the relations existing between users and their preferences, while
they neglect the nature of the items shared in the network and do not exploit
all the available information from STS (i.e., the complete folksonomy).

\textit{Tag-expansion}~\cite{lo2010folksonomy} is the only solution proposed
in the literature to perform folksonomy-based reasoning for content
dissemination in an opportunistic environment. Using this approach, each node
builds a tag co-occurrence matrix to identify the tags that are most frequently
used in conjunction with other tags (i.e., expanded tags), and it downloads an
item if it is tagged with one of these tags. One of the main drawbacks is that a
node could receive more items than those really interesting for it because the
user may not be interested in the topics represented by the expanded tags. In
addition, this approach can suffer from scalability problems depending on the
dimension of the data structures to be kept in memory and on the sparseness of
the matrix. Since the distribution of the popularity of tags in STS generally
follows a long tail distribution~\cite{delicious_stats}, the data representation
in a tag co-occurrence matrix could be often highly sparse.

\begin{figure}[t]
  \centering \includegraphics[width=0.48\textwidth]{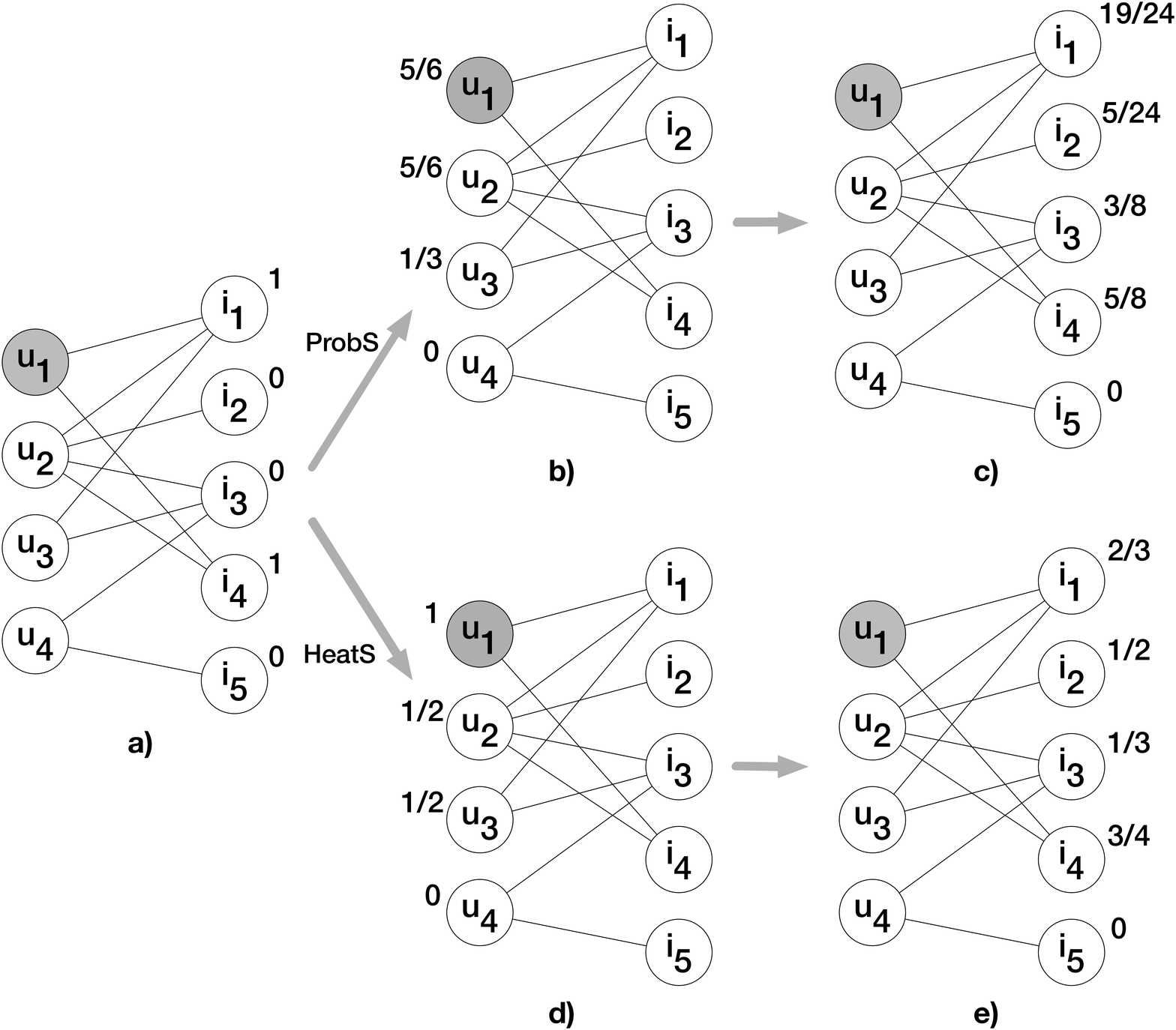}
  \caption[Execution of ProbS and HeatS with a bipartite graph]{Application of
    ProbS and HeatS with a bipartite graph.\cite{zhou2010solving}}
  \label{fig:ex_ProbS_HeatS}
\end{figure}

\subsection{Tag-based Recommender Systems}

To overcome the limitations of tag-expansion in opportunistic networks, we
propose to use a new family of recommender systems based on folksonomies:
\textit{Tag-based Recommender Systems}~\cite{zhang2011tag}. In the literature,
many approaches for Tag-based Recommender Systems have been proposed, but~--~to
the best of our knowledge~--~none of them has been so far employed in
opportunistic networks. Among different possible tag-based solutions,
\textit{diffusion-based}~\cite{zhang2011tag} algorithms are the most promising
ones for our reference scenario. These algorithms try to overcome the complexity
and scalability issues by using graphs as a natural way to represent
folksonomies. In these cases, nodes of these graphs represent users, items
and/or tags, while their links represent relationships among nodes. Since nodes
are divided into three separate classes, folksonomies are usually represented as
tripartite graphs, or by separate bipartite graphs for user-item or user-tag
relationships in order to further reduce the computational complexity. These
approaches rely on the diffusion of fictitious resources within the folksonomy
graph, starting from a node representing a target user (i.e., the target of the
recommendation) and diffusing the resources by following links between
nodes. This permits to identify relevant items (or tags) as those nodes that are
indirectly connected to the target user via other users with whom she shares one
or more connections. The higher the number of links connecting an item $i$ to
the items of the target user, the higher the score that $i$ will receive for the
recommendation. In this way, the recommender system exploits the structure of
the graph to identify content relevant for the user. In addition, this approach
gives the opportunity to exploit additional hidden information derived from
graph-based analysis (e.g., community detection~\cite{fortunato2010community}),
which could be used to further customize recommendations.

\begin{figure}[t]
  \centering \captionsetup[subfloat]{farskip=2pt,captionskip=3pt}
  \subfloat[First recommendations to $U_1$\label{subfig:sem_t1}]{%
    \includegraphics[width=0.48\textwidth]{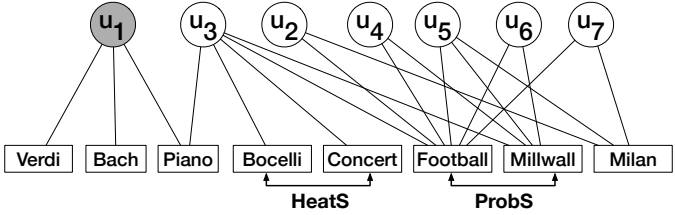} }\\
    \vspace{1.1cm} \subfloat[First recommendations to
      $U_6$\label{subfig:sem_t2}]{%
      \includegraphics[width=0.48\textwidth]{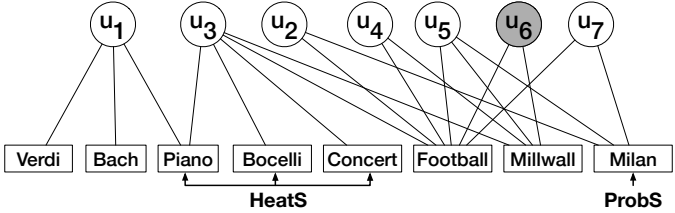} }
    \caption{ProbS and HeatS suggestions.}
    \label{fig:sem_example}
\end{figure}

The oldest and most famous diffusion-based solutions are represented by two
algorithms: \textit{ProbS} (also known as \textit{Mass
  Diffusion})~\cite{zhou2007bipartite, zhang2007recommendation} and
\textit{HeatS}~\cite{zhang2007heat}.  By applying ProbS to user-item bipartite
graphs, a generic resource is assigned to each item $i_s$ directly linked to the
target user $u_t$ of the recommendation such that its value is 1 if an edge is
present between $u_t$ and $i_s$, and 0 otherwise.  These items represent the
content that the user has created or downloaded in the past.  The resource is
then split (\textit{first diffusion step}) among the users directly connected to
the item and each user receives the same portion of the resource. Subsequently,
each user splits the portion of the received resource among the items connected
to her (\textit{second diffusion step}) and each item then receives the same
portion of the resource. The final score of each item $i_j$ is given by the sum
of the portions of resources that are assigned to it after the two diffusion
steps. The set of all the scores obtained in this way is called \textit{resource
  vector} and it can be used to rank the items not directly linked to the target
user. The higher the score obtained by an item, the greater could be the
interest in it for the target user.  The mechanism of HeatS is similar to ProbS,
but it is based on opposite rules: each time a resource (or a portion of it) is
redistributed, it is divided by the number of edges connected to the node
towards which it is heading to. Figure \ref{fig:ex_ProbS_HeatS} depicts the
diffusion steps of the two algorithms, highlighting the differences in the two
recommendations.  Although they may seem a good way to make recommendations,
actually both of them are biased by the presence of extremely popular or
non-popular items or tags, and they do not take into account the characteristics
of the user's interests. For this reason, they present strong limitations if
applied to STS. Specifically, ProbS tends to recommend most popular contents,
while HeatS tends to highlight those with minimal popularity (i.e., with the
smallest possible number of users connected to them). Figure
\ref{fig:sem_example} depicts an example of the described behavior. If we
consider a target user interested in tags with low popularity ($u_1$ in Figure
\ref{subfig:sem_t1}), HeatS will suggests tags with low popularity (that,
however, may be of limited interest for the target user from a semantic point of
view). On the other hand, ProbS tends to recommend tags with high popularity,
possibly semantically unrelated with the user's interests. By contrast, if we
consider a user with popular interests ($u_6$ in Figure \ref{subfig:sem_t2}),
ProbS highlights the correct tags and, instead, HeatS still (wrongly) recommends
the less popular tags.

To overcome these limitations, a \textit{ProbS+HeatS} hybrid approach (hereafter
just \textit{Hybrid})~\cite{zhou2010solving} has been recently proposed in the
literature. This algorithm calculates a linear combination of the results of
ProbS and HeatS with a parameter $\lambda$ governing the relative importance of
one of the two original algorithms. However, the problem of Hybrid (and other
recently proposed solutions~\cite{lu2011information, liu2011information,
  zeng2014uncovering, ma2016personalized}), precisely lies in the use of
parameters which can vary greatly depending on the nature of the dataset, and
that are difficult to estimate in real situations.

PLIERS~\cite{arnaboldi2016pliers} solves the dilemma of the choice between
popular or non-popular items in the network in a more natural way than the other
diffusion-based algorithms, without requiring any parameters to tune, and
ensuring that the popularity of recommended items is always comparable with the
popularity of items already adopted by the users. PLIERS assumes that if the
user is interested in general categories of items, with a high number of
connected users (i.e., popular items), the recommendations will be general as
well. On the other hand, if the user is interested in less popular items, the
recommendations will prefer items with less connected users.

\begin{figure}[t]
  \centering
  \includegraphics[width=0.48\textwidth]{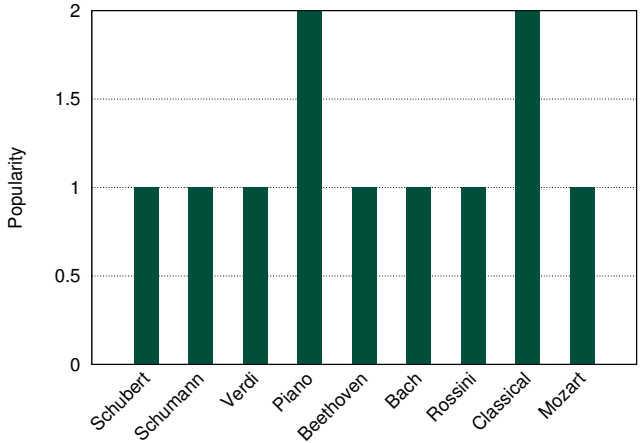}
  \caption{Tags popularity for User 1.}
  \label{fig:user_1_tags_popularity}
\end{figure}

In addition, PLIERS assumes that a very popular item/tag can semantically relate
to a more ``generic'' topic compared to a less popular item/tag that, instead,
describes a more ``specific'' topic. For example, any content related to the
football club Millwall can be tagged with both tags \textit{``Millwall''} and
\textit{``Football''} but the opposite is not always true: all content
concerning football will not always be tagged with
\textit{``Millwall''}. According to this assumption, we can therefore say that
the tag \textit{``Football''} refers to a more generic topic than that referred
by the tag \textit{``Millwall''}. Users interested in the Millwall football
club, but not connected to items tagged with \textit{``Football''}, are clearly
not interested in all the items tagged with the latter tag, as these could
contain information about other football clubs. PLIERS leverages this assumption
to improve the recommendations and to provide more personalized items to the
users. We demonstrated that PLIERS outperforms all the other diffusion-based
approaches applied to Online Social Networks, both in terms of recommendation
accuracy and personalization~\cite{arnaboldi2016pliers}. In this paper we
present p-PLIERS, a framework able to merge the knowledge about users, interests
and contents on single mobile devices and to evaluate the relevance of the
available contents for each single user through PLIERS recommendations. To make
the reader able to completely understand this new solution, we provide, in the
next sections, an extended description of PLIERS notation and algorithm with
respect to~\cite{arnaboldi2016pliers}. Then, in Section~\ref{the_algorithm}, we
present p-PLIERS.

\begin{table}[t]
\centering
\caption{Ranking for User 1 for PLIERS (PL), ProbS (Pr), HeatS (He) and Hybrid
  (Hy).}
\label{tab:user1_rank}
\begin{tabular}{lrrrr}
    \toprule {\bf Tag} & {\bf PL} & {\bf Pr} & {\bf He} & {\bf Hy} \\ \midrule
    orchestra (1) & 1 & 12 & 1 & 1 \\ debussy (1) & 2 & 13 & 2 & 2
    \\ thepianoguys (1) & 3 & 16 & 3 & 3 \\ pavarotti (1) & 4 & 19 & 4 & 4
    \\ nabucco (1) & 5 & 20 & 5 & 5 \\ millwall (22) & 13 & 5 & 14 & 13 \\ music
    (24) & 15 & 4 & 15 & 14 \\ sherlock (27) & 16 & 3 & 16 & 15 \\ sport (41) &
    19 & 2 & 19 & 18 \\ startrek (44) & 22 & 1 & 22 & 19 \\ \bottomrule
\end{tabular}
\end{table}

\section{PLIERS: PopuLarity based ItEm Recommender System}
\label{pliers}
In this section, we present PLIERS working principles on a simple bipartite
graph, and its comparison with ProbS, HeatS and Hybrid. Then, we describe in
detail how PLIERS can be applied to a tripartite graph, which is used by
p-PLIERS to represent knowledge in opportunistic scenarios.

\subsection{Notation}
\label{sec:pliers_notation}
Formally, a folksonomy can be represented with three node sets: users $U =
\{u_1, \ldots , u_n\}$, items $I = \{i_1, \ldots, i_m\}$ and tags $T = \{t_1,
\ldots, t_k\}$.  Consequently, each binary relation between them can be
described using adjacency matrices, \textbf{$A^{UI}$}, \textbf{$A^{IT}$},
\textbf{$A^{UT}$} respectively for user-item, item-tag and user-tag
relations. If the user $u_l$ has collected the item $i_s$, we set $a^{UI}_{l,s}
= 1$, otherwise $a^{UI}_{l,s} = 0$. Similarly, we set $a^{IT}_{s,q} = 1$ if
$i_s$ has been tagged with $t_q$ and $a^{IT}_{s,q} = 0$ otherwise. Furthermore,
connections between users and tags can be represented by an adjacency matrix
\textbf{$A^{UT}$}, where $a^{UT}_{l,q} = 1$ if $u_l$ owns items tagged with
$t_q$, and $a^{UT}_{l,q} = 0$ otherwise.

In the next subsection, we consider user-item bipartite graphs described by the
$A^{UI}$ adjacency matrix.

\subsection{The algorithm}
\label{sec:pliers_the_algo}

PLIERS is inspired by ProbS and shares with it the same two diffusion steps. In
addition, PLIERS normalizes the value obtained by ProbS when comparing an item
$i_j$ with one of the items $i_s$ of the target user $u_t$. This normalization
is performed by multiplying the recommendation score by the cardinality of the
intersection between the set of users connected to $i_j$ and the set of users
connected to $i_s$, divided by $k(i_j)$, which is the popularity of $i_j$. In
this way, items with popularity similar to the popularity of the items of the
target user, and that possibly share the same set of users, are preferred.

The final value of the item $i_j$ for the target user $u_t$, in a graph with $n$
users and $m$ items, is then calculated by PLIERS with the following formula.

\begin{equation}
  \vspace{-0.1cm} \small f^{pl}_j = \sum_{l = 1}^{n}\sum_{s = 1}^{m} \frac{a_{l,j}
    \cdot a_{l,s} \cdot a_{t,s}}{k(u_l) \cdot k(i_s)} \frac{\left | U_s \cap U_j
    \right |}{k(i_j)} \hspace{0.3cm} j = 1,\ldots,m ,
\label{eq:pliers_formula}
\end{equation}
Here, $U_j$ is the set of users connected to the item $i_j$, $k(i_j)$ is the
popularity degree of the item $i_j$ (i.e., the number of connected users), and
$a_{x,y}$ is an element of the $A^{UI}$ matrix.  The higher the value of
$f^{pl}_j$, the more item $i_j$ is similar to the items already owned by $u_t$.

\subsection{Working principles of PLIERS on a simple graph}

To describe in detail the mechanism of PLIERS, and to prove its effectiveness,
we manually built a synthetic user-tag bipartite graph consisting of 64 users,
62 tags and a total of 612 edges. The graph represents the relation between a
set of users and several tags, which are characterized by a variable degree of
popularity. A link between a user and a tag indicates that the user owns that
tag (i.e., she has already downloaded a content labeled with that tag).

\begin{figure}[t]
    \centering
    \includegraphics[width=0.48\textwidth]{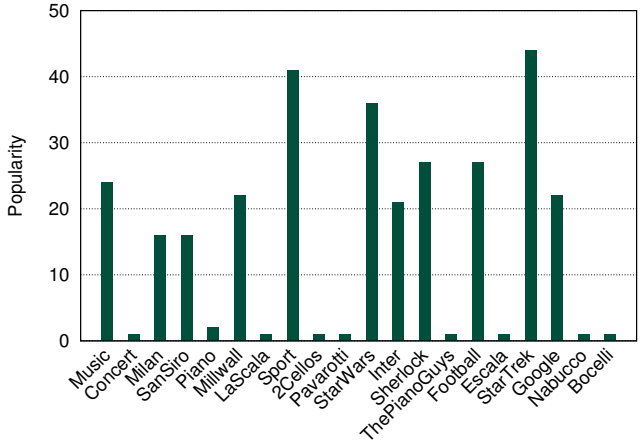}
    \caption{Tags popularity for User 3.}
    \label{fig:user_3_tags_popularity}
\end{figure}

\begin{figure}[b]
  \centering
  \includegraphics[width=0.48\textwidth]{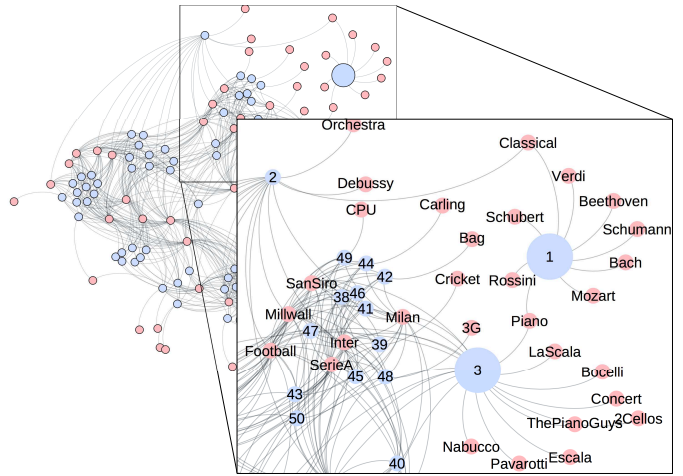}
  \caption{Structure of the synthetic user-tag bipartite graph.}
  \label{fig:pliers_example_graph}
\end{figure}

Figure~\ref{fig:pliers_example_graph} shows the structure of a synthetic
bipartite graph, highlighting the characteristics of two opposite cases: the
first user (User 1), connected just to non-popular tags (more specific), and the
second one (User 3), connected to tags with, on average, higher popularity. To
clarify the difference between popular and non-popular tags: the tag
``TV-Series'' is far more popular than ``SherlockHolmes'' and, since the
majority of users connected to the tag ``SherlockHolmes'' are also connected to
``TV-Series'' (but the opposite is not always true), semantically speaking, we
can refer to ``TV-Series'' as a ``superclass'' of ``SherlockHolmes''.

\begin{table}[t]
  \centering
  \caption{Ranking for User 3 for PLIERS (PL), ProbS (Pr), HeatS (He) and Hybrid
    (Hy).}
  \label{tab:user_3_rank}
  \small
  \begin{tabular}{@{}lrrrr@{}}
    \toprule {\bf Tag} & {\bf PL} & {\bf Pr} & {\bf He} & {\bf Hy} \\ \midrule
    3g (1) & 26 & 26 & 5 & 18 \\ bag (1) & 24 & 24 & 4 & 17 \\ carling (1) & 23
    & 23 & 3 & 16 \\ cpu (1) & 22 & 22 & 2 & 15 \\ cricket (1) & 25 & 25 & 1 &
    14 \\ seriea (19) & 1 & 1 & 6 & 1 \\ android (21) & 4 & 8 & 21 & 6
    \\ googleglass (21) & 3 & 7 & 20 & 5 \\ mobile (21) & 2 & 6 & 19 & 4
    \\ crime (23) & 12 & 4 & 23 & 7 \\ pop (23) & 8 & 5 & 24 & 8 \\ tv-series
    (34) & 9 & 3 & 25 & 3 \\ sci-fi (37) & 5 & 2 & 22 & 2 \\ \bottomrule
  \end{tabular}
  \end{table}

In the following, we compare the differences in the recommendations lists (i.e.,
the resource vectors) generated by PLIERS, ProbS, HeatS and Hybrid for the two
considered users.

\textbf{User 1} is interested in topics related to classical music
(``Schubert'', ``Verdi'', ``Rossini'', etc.) that, in our graph, have a low
popularity with mean equal to 1.22. Figure~\ref{fig:user_1_tags_popularity}
depicts the popularity of each tag connected to User 1 in terms of number of
users connected to that tag.

Table~\ref{tab:user1_rank} contains the results obtained from the execution of
the four algorithms, considering User 1 as the target. We reported the top 10
tags common to the rankings of all the algorithms, together with their position
in each ranking, so as to better compare the differences among the approaches.
Next to each tag, its popularity is reported.  PLIERS, HeatS and Hybrid (with
$\lambda=0.5$~--~the value generally used when there is no a priori knowledge on
the data) recommend similar tags for the first 12 positions, highlighting tags
with a popularity degree similar to those already connected to the user, and
more semantically related with them (e.g., ``Debussy'', ``Orchestra''). On the
contrary, ProbS presents wrong recommendations in the first positions, by
assigning a higher score to uncorrelated tags (from a semantic point of view),
which are characterized with a popularity degree that deviates too much from
that of the tags held by the user (e.g., ``StarTrek'', ``Sport'').

By contrast, \textbf{User 3} is interested in both non-popular tags (such as
``Pavarotti'', ``2Cellos'', and ``Nabucco'') and particularly popular
(``StarTrek'', ``Star Wars'', ``Millwall'') or generic tags (``Sport'',
``Music''), which increase the average popularity of her topics to the value of
15.3 (Figure~\ref{fig:user_3_tags_popularity}). In this case, PLIERS adapts its
recommendations to the ``generic'' nature of the target's interests, deviating
from HeatS and agreeing with the suggestions made by ProbS and Hybrid (see
Table~\ref{tab:user_3_rank}).

\subsection{Extension to tripartite graphs}

The three adjacency matrices introduced in Section \ref{sec:pliers_notation} can
be represented as a tripartite graph $G = (U,I,T,E,F)$ where $U$ is the set of
users in the folksonomy, $I$ is the set of items, $T$ is the set of tags, $E$ is
the set of links between users and items, and $F$ is the set of links between
items and tags. If an edge $e_{l,s}$ between the user-node $u_l$ and the
item-node $i_s$ exists, we say that the user $u_l$ was interested in the item
$i_s$ (i.e., she created or downloaded it in the past) and, in a completely
analogous way, if the item-node $i_s$ is connected to the tag-nodes
$t_i,\ldots,t_k$, we mean that the item $i_s$ was tagged with $t_i,\ldots,t_k$.

Recommender systems based on tripartite rather than bipartite graphs exploit all
the information available from social tagging systems, and this can lead to
better results in terms of recommendation accuracy and precision, as we show in
Section \ref{sec:experiments_static}.

\begin{figure}[t]
  \center
  \includegraphics[width=0.47\textwidth]{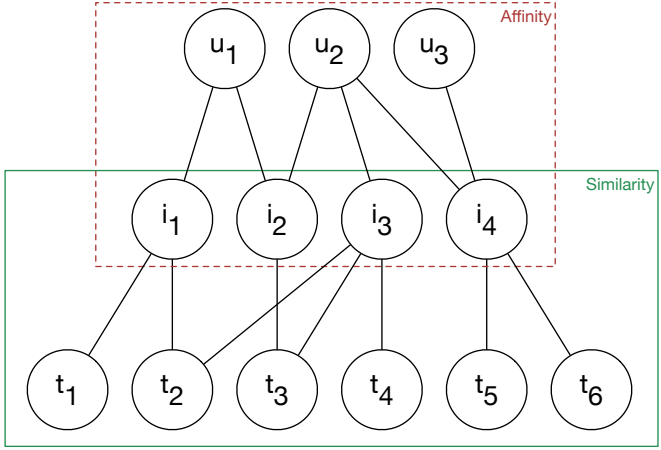}
  \caption{Affinity and similarity indices in a tripartite folksonomy graph.}
  \label{fig:affinity_similarity}
\end{figure}

As depicted in Figure~\ref{fig:affinity_similarity}, in a tripartite graph we
can define two characteristic values:

\begin{description}
	\item[Affinity index:] is the score obtained by PLIERS when applied on
          user-item links. This indicates the degree to which an item is close
          to the user's interests and preferences.

	\item[Similarity index:] is the score obtained by PLIERS when applied on
          tag-item links. This indicates the degree to which two items are
          similar in terms of the tags with which they are labeled.
\end{description}

More formally, we define the \textit{affinity} index of an item $i_j$ for a
target user $u_t$ with the following formula:

\begin{equation}
  \vspace{-0.1cm} \small f^a_j = \sum_{l = 1}^{n}\sum_{s = 1}^{m} \frac{a_{l,j}
  \cdot a_{l,s} \cdot a_{t,s}}{k_i(u_l) \cdot k_u(i_s)} \cdot \frac{\left | U_s
  \cap U_j \right |}{k_u(i_j)},
\label{eq:affinity_index}
\end{equation}
where $U_j$ is the set of users connected to the item $i_j$, $k_u(i_j)$ is the
popularity degree of the item $i_j$ (i.e., the number of connected users),
$k_i(u_l)$ is the number of items connected to the user $u_l$, and $a_{x,y}$ is
an element of the matrix $A^{UI}$.

Symmetrically, we define the \textit{similarity} index of an item $i_j$ with
respect to the items already linked to the target user $u_t$, as

\begin{equation}
  \vspace{-0.1cm} \small f^s_j = \sum_{z = 1}^{k}\sum_{s = 1}^{m} \frac{a_{z,j}
  \cdot a_{z,s} \cdot a_{t,s}}{k_i(t_z) \cdot k_t(i_s)} \cdot \frac{\left | T_s
  \cap T_j \right |}{k_t(i_j)},
\label{eq:similarity_index}
\end{equation}
where $T_j$ is the set of tags connected to the item $i_j$, $k_t(i_j)$ is the
number of tags with which the item $i_j$ was marked, $k_i(t_z)$ is the number of
items connected to the tag $t_z$, and $a_{x,y}$ is an element of the matrix
$A^{IT}$.

We define the final score of an item $i_j$ as the linear combination of the two
indices (\ref{eq:affinity_index}) and (\ref{eq:similarity_index}) as follows.

\begin{equation}
  \vspace{-0.1cm} \small f_j = \lambda \cdot f^a_j + (1 - \lambda) \cdot f^s_j,
\label{eq:final_score_tripartite}
\end{equation}
where $\lambda \in [0,1]$ is a tunable parameter of the algorithm with which we
can weigh the links between users and items and those between items and tags.

\section{Pervasive PLIERS}
\label{the_algorithm}

p-PLIERS implements a framework for the representation and exchange of context
information describing contents and users among nodes of an opportunistic
network. It exploits PLIERS for the evaluation of the collected knowledge and to
provide personalized recommendations to the users about available interesting
contents.

\begin{algorithm}[b]
  \caption{p-PLIERS: Content discovery and evaluation in opportunistic
    networks}
	\label{opp_algorithm}
	\begin{algorithmic}[1]
		\Procedure{Encounter}{node $n$}\
    \State Send my $LKG$ to $n$
    \State $n_{LKG} \gets$ Receive $n$'s $LKG$
    \State Update my $LKG$ with $n_{LKG}$
    \State $I \gets$ new discovered items
		\For{each item $i \in I$} \State $score(i) \gets $ evaluate $i$
                with PLIERS \EndFor \EndProcedure
	\end{algorithmic}
\end{algorithm}

\begin{figure*}[t]
  \centering \captionsetup[subfloat]{farskip=2pt,captionskip=3pt}
  \subfloat[\label{fig:expo_user_tweets_static}]{%
    \includegraphics[width=0.33\textwidth]{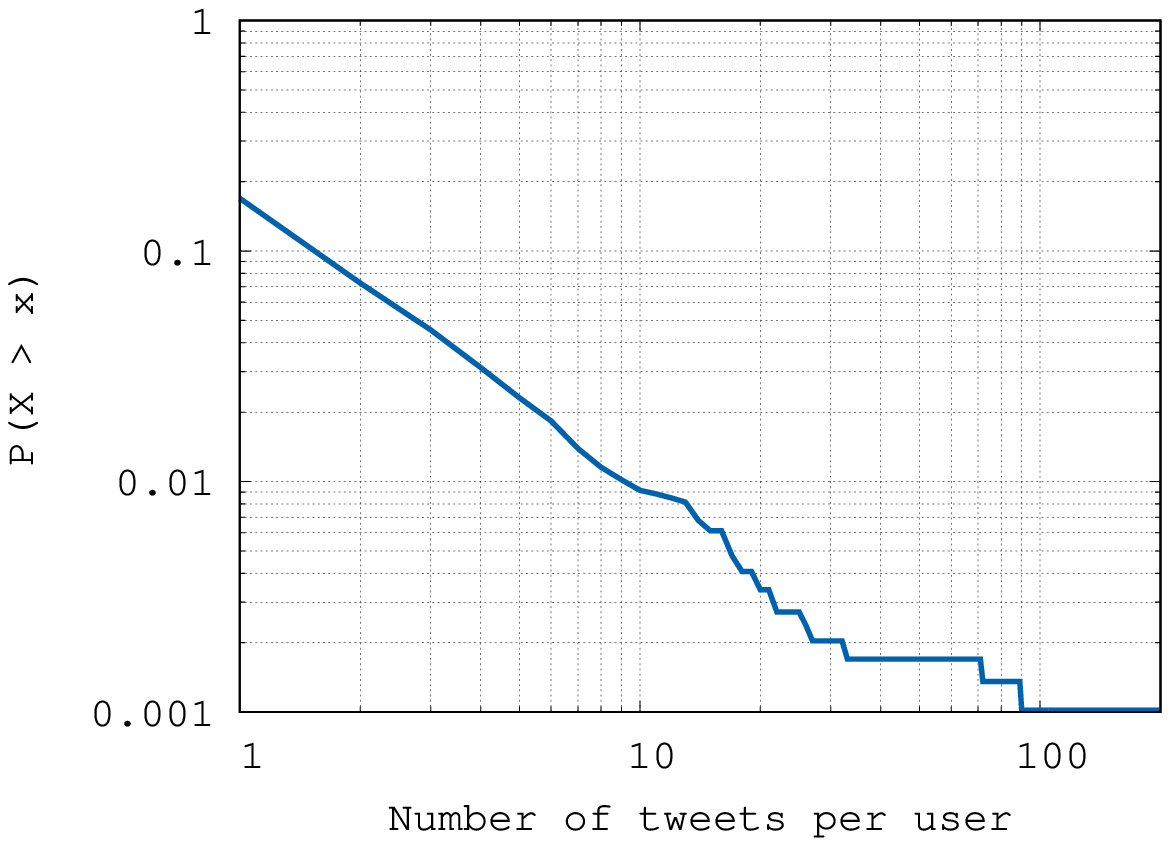} }
  \subfloat[\label{fig:expo_tweet_tags_static}]{%
    \includegraphics[width=0.33\textwidth]{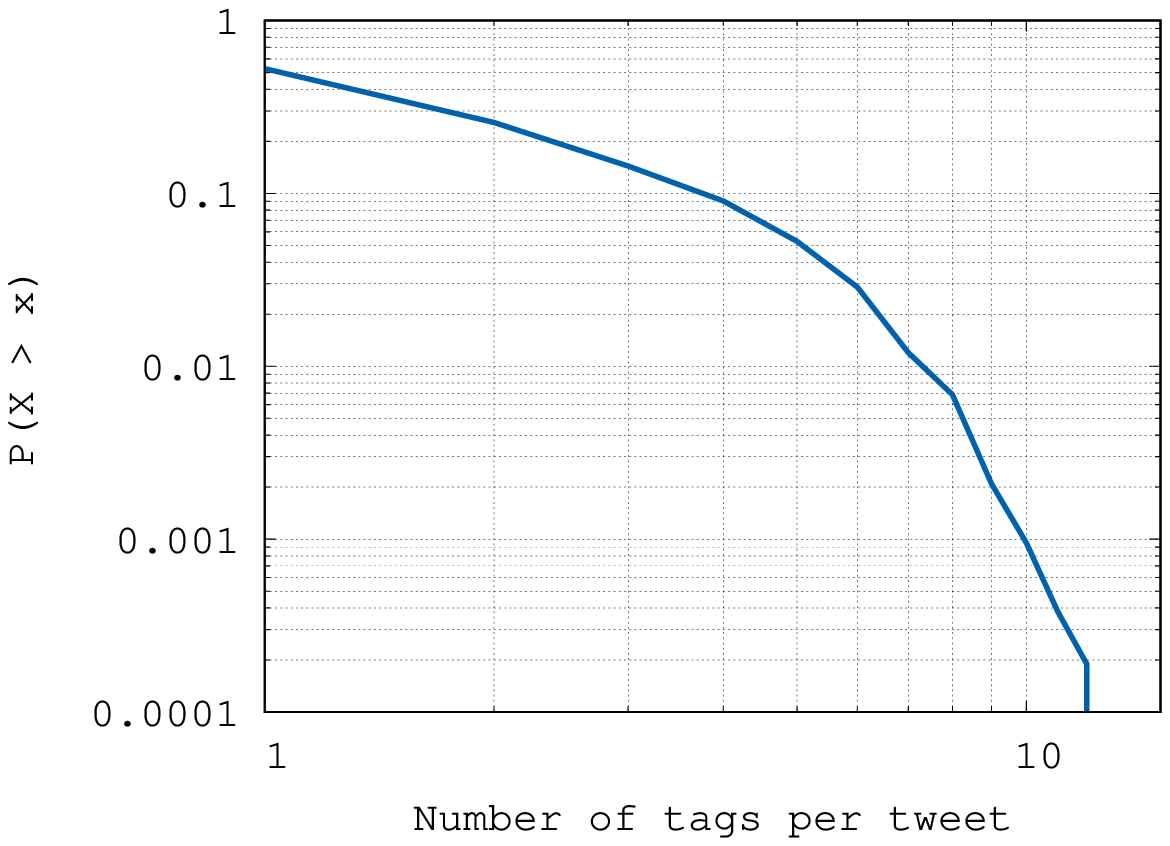} }
  \subfloat[\label{fig:expo_most_used_tags_static}]{%
    \includegraphics[width=0.33\textwidth]{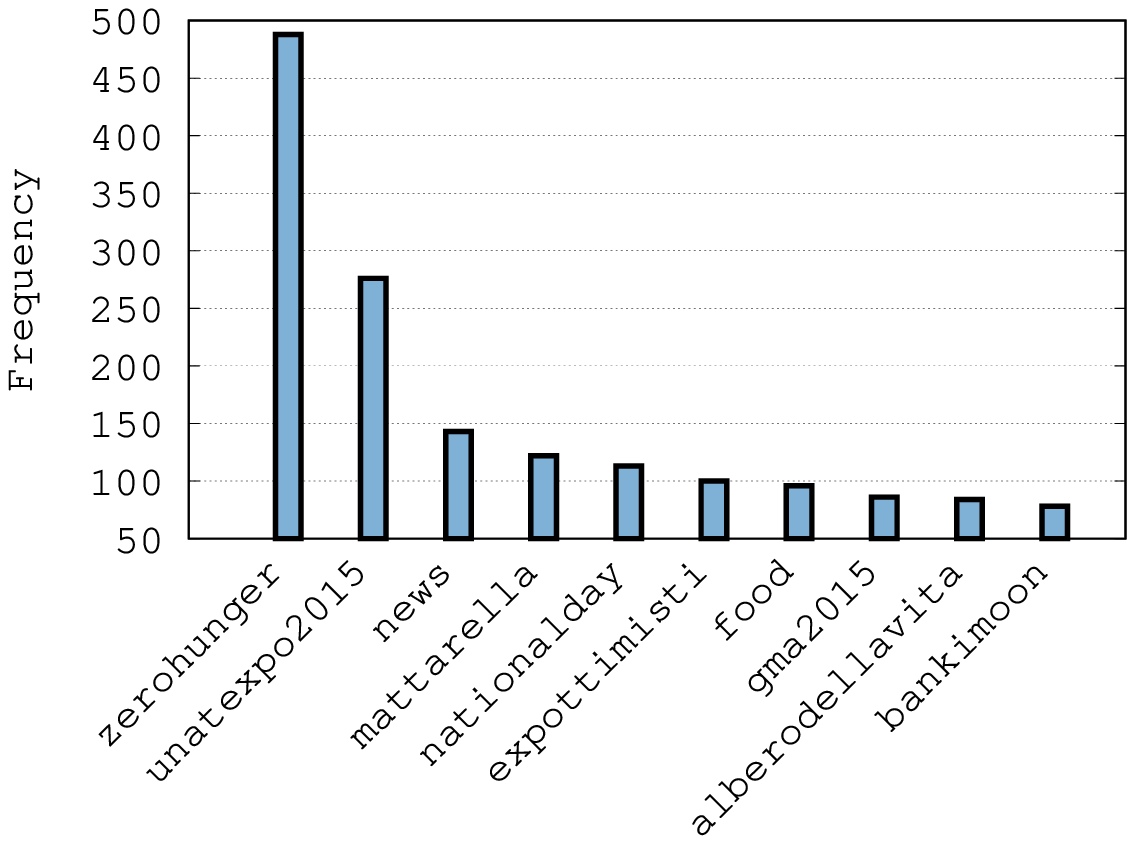} }
    \caption{Descriptive statistics of the Twitter dataset used for the
      WFD@Expo2015 scenario. (a) CCDF of the number of tweets per user, (b) CCDF
      of the number of tags per tweet, and (c) Frequency of use of the most
      used tags.}
    \label{fig:expo_stats}
\end{figure*}

To this aim, p-PLIERS supports the maintenance of a \emph{local} representation
of the knowledge about the users, items and tags in the network and their
relations (i.e., a local knowledge graph - LKG) on each node of the
network. This knowledge is built by merging the information about contents
created or downloaded by each local user, with the local knowledge of other
nodes gathered during physical contacts. The nodes use this partial
knowledge to evaluate the relevance of items (content) carried by other nodes in
proximity, with respect to the interests of their local user. The evaluation is
performed locally, without requiring centralized control and without the need of
having a \emph{global} knowledge of the network (i.e., global knowledge graph -
GKG).  We represent the LKG of each node as a tripartite graph as described in
the previous section.
When a node creates a new item, it updates its LKG by inserting the relation
between the user entity that represents it in the graph and the created item,
and the relations between the item and its tags. When a node encounters another
node in the network, the two nodes exchange their LKGs, and locally integrate
them with the received information. The basic operations of our solution are
summarized in Algorithm~\ref{opp_algorithm}.

Our solution relies on the fact that, as we will demonstrate with simulations,
using the local knowledge of each node is sufficient to correctly evaluate the
relevance of contents, and leads to recommendations that are compatible with the
results that one would obtain with a \emph{global} knowledge about all the
information available in the whole network at the time of the evaluation. This
supports the decentralization of recommendations, and allows us to efficiently
apply recommender systems, and PLIERS in particular, to dynamic mobile
environments.

\section{Experimental Evaluation: General Description}
\label{sec:experiments_gen}

As a first set of experiments, we evaluated the accuracy of PLIERS
recommendations with respect to other recommender systems typically used for
content dissemination in opportunistic networks. To do so, we considered a
\emph{static} user-item-tag graph obtained from Twitter to evaluate the accuracy
of the recommendations given by PLIERS. Then, we thoroughly evaluated p-PLIERS
in three different \emph{dynamic} opportunistic scenarios. In both cases (static
and dynamic), we used graphs derived from real online tagged contents obtained
from Twitter, and we set the parameter $\lambda$ in equation
\ref{eq:final_score_tripartite} to 0.5.

\begin{table}[b]
  \caption{Statistics of the Twitter dataset used for the simulations.}
  \label{tab:twitter_stats}
  \centering
  \begin{tabular}{l  r}
    \hline \textbf{Statistic} & \textbf{Value}\\ \hline N. of tweets & 5,260
    \\ N. of users & 2,946\\ N. of unique hashtags & 3,292\\ Average n. of
    hashtags per tweet & 2.12\\ Max n. of hashtags per tweet & 13\\ Min n. of
    hashtags per tweet & 1\\ \hline
  \end{tabular}
\end{table}

The static scenario allowed us to evaluate the accuracy of PLIERS using a
standard evaluation method (i.e. \emph{link prediction}, as detailed in the
following). For this scenario, we downloaded the tweets generated during World
Food Day at Expo 2015 (WFD@Expo2015) in the urban area of Milan, and we built a
tripartite graph composed by users, tweets, and their hashtags. This represents
a realistic folksonomy graph of online tagged contents related to a popular
event.

For the dynamic scenarios, we considered different situations in which pervasive
communication systems may be used. Specifically, the scenarios are characterized
by variable number of people moving with different mobility patterns generated
by human mobility models or derived from real contact traces. Each scenario
represents a specific application use case: (i) a big event (WFD@Expo2015), (ii)
a conference (ACM KDD'15), and (iii) an urban area (Helsinki city center). We
performed a set of simulations based on these scenarios, where each person is
expected to use a mobile device able to communicate through D2D communications
with other devices in proximity. In addition, each device allows the users to
generate and tag contents over time, and uses p-PLIERS to identify potentially
interesting contents created by others. We think that the scenarios cover a
significant set of cases where pervasive and mobile systems may be applied, and
represent thus the basis for realistic experimental evaluations.

\section{PLIERS Experimental Evaluation in a Static Scenario}
\label{sec:experiments_static}

We compared the performances of PLIERS with respect to the other recommender
systems proposed in the literature for opportunistic networks, namely user-based
\emph{Collaborative Filtering} and \emph{Tag Expansion}. To assess the accuracy
of the recommendations, we performed a \textit{link prediction task} on a
tripartite graph. This is a standard way to evaluate and compare recommender
systems~\cite{kantor2011recommender}. In essence, the technique consists in
randomly removing a small portion of links from a folksonomy graph, then
verifying whether the recommendations generated on the pruned graph coincide
with the removed links. Intuitively, a good recommender system should be able to
identify the items that were originally connected to the removed links.

\subsection{Dataset Description}

We downloaded a dataset of 5,260 tweets generated during World Food Day at Expo
2015 through Twitter Streaming API. To do so, we applied a filter to Twitter API
by indicating a set of hashtags (e.g., \#Expo2015, \#ExpoMilan2015, and other
possible combinations) and we queried only tweets generated in the area of
Milan. We decided to validate PLIERS using tweets generated in a limited area
during a thematic event such as the WFD@Expo2015 since we think that the
semantic relationships between the relative folksonomy entities is significant,
and we expect to obtain meaningful and realistic recommendations.

The number of different users that generated the downloaded tweets is 2,946. We
removed the hashtags that we used to filter the data (e.g., \#Expo2015,
\#ExpoMilan) as they are contained in all the tweets, and they are thus not
useful to describe the contents. In Table~\ref{tab:twitter_stats}, we report the
statistics of the dataset. Figure~\ref{fig:expo_tweet_tags_static} and
Figure~\ref{fig:expo_user_tweets_static} depict the CCDF of the number of tags
per tweet and the number of tweets per user. It is worth noting that both CCDFs
show a long-tailed distribution, a typical result for social networks. In
addition, in Figure~\ref{fig:expo_most_used_tags_static}, we depict the
frequency of use of the 10 most used tags in the dataset.

\begin{figure}[t]
  \center \includegraphics[width=0.45\textwidth]{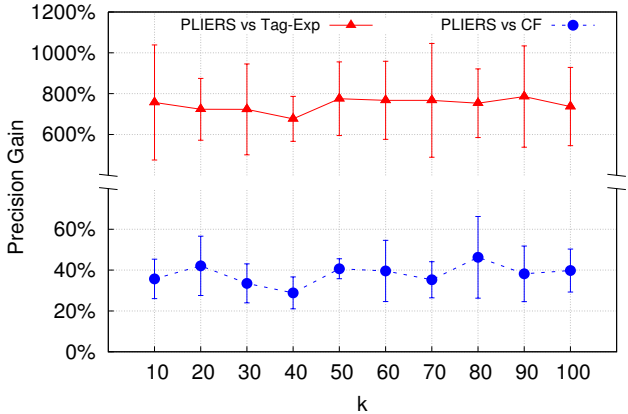}
  \caption{Precision gain obtained by PLIERS in a complete knowledge scenario.}
  \label{fig:precision_complete}
\end{figure}

\subsection{Link Prediction Task}

We removed 1 link from each user connected to at least 5 items with popularity
greater than 1, and then we ran PLIERS and the other systems on the updated
graph to generate the recommendation list for each user. Then, we calculated the
percentage of removed links that are included in the recommendations of each
algorithm (i.e., the ``recovered links''). We selected the links to be removed
in this particular way in order to avoid the complete isolation of the items
connected to just a single user. In fact, in those cases, all the recommender
systems would not be able to recover the removed links, reducing the
significance of the experimental results. The percentage of links between users
and items we deleted from the original graph is equal to $1.4\%$. We computed
the performances of each method using the measures of Precision (P) and Recall
(R)~\cite{kantor2011recommender}, defined as follows:

\begin{equation}
	P = \frac{1}{\left | U \right |} \sum_{u \in U} \frac{1}{\left | T(u)
          \right |} \sum_{t \in T(u)} \frac{1}{pos(t)},
\label{eq:precision}
\end{equation}

\begin{equation}
	R = \frac{1}{\left | U \right |} \sum_{u \in U} \frac{\left | L(u) \cap
          T(u) \right |}{\left | T(u) \right |},
\label{eq:recall}
\end{equation}
where $U$ is the set of users for whom we have removed links, $L(u)$ is the
recommendation list for the user $u$, $T(u)$ is the set of links removed from
the user $u$, and $pos(t)$ is the position in which the removed item $t$ appears
in the list of recommended items $L(u)$.

\subsection{Results}

\begin{figure}[t]
  \center \includegraphics[width=0.45\textwidth]{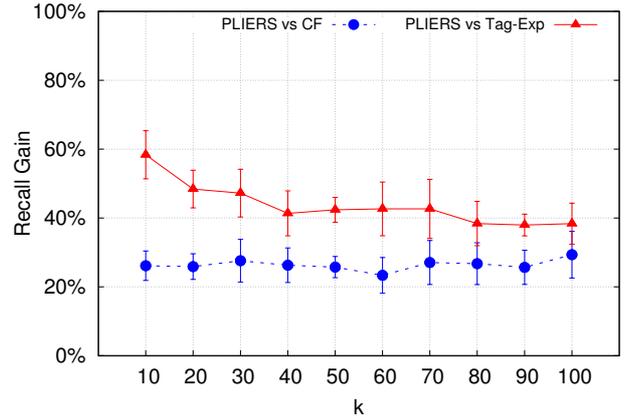}
  \caption{Recall gain obtained by PLIERS in a complete knowledge scenario.}
  \label{fig:recall_complete}
\end{figure}

Figures \ref{fig:precision_complete} and \ref{fig:recall_complete} depict
respectively the Precision gain and the Recall gain of PLIERS with respect to
both CF and Tag Expansion. The gains are expressed in percentage and identify
the improvement obtained by PLIERS in terms of precision and recall with respect
to the other algorithms. The parameter $k$ in the figures represents, for CF, the
number of users similar to the target user to be considered in the calculation,
and for Tag Expansion the $k$ tags more correlated with the tags of the target
user to be considered. Higher values of $k$ should therefore lead to better
recommendations. It is worth noting that PLIERS outperforms the two reference
algorithms for both measures, receiving a precision score up to eight times
higher than that obtained by Tag Expansion and a score 40\% higher than CF, even
in case of very high values of $k$. With regard to the Recall measure, PLIERS
obtains a score around 50\% higher than that received by Tag Expansion and
around 30\% higher than that of CF.

These results indicate that PLIERS, being able to exploit all the information
contained in the folksonomy graph (i.e., user-item-tag relationships), obtains
better results than the state-of-the-art solutions used in opportunistic
networks, which consider only partial user-item or item-tag relationships.

\section{p-PLIERS Experimental Evaluation in Dynamic Scenarios}
\label{sec:experiments_dynamic}

In~\cite{elsherief2015information}, ElSherief et al. established theoretical
limits on the performance of knowledge sharing in opportunistic social
networks. They calculated how many contacts are needed to ensure that nodes are
able to well approximate the global knowledge (i.e., all the available
information in the network) for different sharing policies in a scenario where
the global knowledge is static and defined a priori.

We performed an empirical evaluation similar to that carried out
in~\cite{elsherief2015information}, but using more complex and realistic
reference scenarios, where the information is dynamically generated over time by
the nodes. Specifically, in our reference scenarios, the users are characterized
by different mobility and content generation patterns.

\subsection{p-PLIERS Pervasive Simulator}

To simulate the dynamic scenarios, we implemented a high-level simulator that
emulates the execution of p-PLIERS on a set of mobile nodes that generate
contents over time. The simulator requires in input (i) a set of traces defining
the physical contacts between nodes over time and (ii) a list of contents
generated by the nodes and marked with a timestamp. Then, it calculates
statistics to evaluate p-PLIERS both in terms of its ability to approximate the
global knowledge from a local perspective, and the accuracy of the given
recommendations. The simulation process proceeds in discrete steps, each of
which represents 1 minute of simulated time. For each contact in the simulation
steps, indicated by the timestamps in the contact traces, the involved nodes
establish a D2D communication between each other and they perform the
operations of Algorithm~\ref{opp_algorithm}.

Note that the simulator is at a higher abstraction level than other existing
network simulators (e.g., generic network simulators such as
ns-3\footnote{https://www.nsnam.org/} or OMNeT++\footnote{https://omnetpp.org/},
and simulators specific for opportunistic networks such as
TheONE\footnote{https://akeranen.github.io/the-one}). Thus, at this level, we do
not consider network-related issues, but we assume that, when two nodes are in
contact, the communication channel between them was successfully established.

To represent the knowledge about the available content in the simulated mobile
network, each node $u$ maintains a \emph{local} tripartite graph $LKG_u$ that
represents its (partial) knowledge about the associations between the contents
in the network and the nodes which created them, and the associations between
the contents and the tags associated with them.

The $LKG$ of each node is initially empty. Every time a new item (tweet) $i$ is
created by node $u$, $u$, $i$, and each tag $t$ associated with $i$ are added to
$LKG_u$ (if not present), as well as the link connecting $u$ and $i$ ($e_{ui}$)
and all the links connecting $i$ and each of its tags $t$ ($f_{it}$).

We also maintain $GKG$, the \emph{global} tripartite graph that represents the
complete knowledge of all the users, items, and tags in the network at a certain
time.

\subsection{Measures}

At each time step of the simulation (i.e., every minute), the simulator
calculates the following measures:

\begin{enumerate}
\item Average similarity between the $LKG$ of each node and the $GKG$.
\item Average similarity between the vector of recommendations generated by
  PLIERS on the local and global graphs.
\end{enumerate}

To calculate the similarity between tripartite graphs, we first flattened the
graphs, obtaining two adjacency lists $L_1$ and $L_2$, and then calculated the
Jaccard index $J$ on these lists:

\begin{equation}
  J(L_1,L2) = \frac{|L_1 \cap L2|}{|L_1 \cup L2|}
\end{equation}

In the same way, the similarity of two ranked recommendation vectors $R_1$ and
$R_2$ is calculated as $J(R_1,R_2)$.

Although the Jaccard index is a standard and widely used measure of similarity,
it does not consider possible differences in terms of position in the rankings
given by the recommendation vectors, but rather considers only the presence or
absence of recommended items in the two vectors. To account for possible
differences in terms of position of the same elements in the rankings produced
either using local or global knowledge graphs, we also calculated a similarity
measure based on Spearman's Footrule~\cite{brandenburg2013nearest} defined as
follows:

\begin{equation}
  S(R_1, R_2) = 1 - \frac{\sum_{x \in R_1 \cap R_2}{d(x, R_1,
      L_2)}}{max(|R_1|,|R_2|)},
\end{equation}

where

\begin{equation}
  d(x, R_1, R_2) =
  \begin{cases}
    |R_1(x) - R_2(x)|& \text{if $x \in R_1 \cap R_2$}\\ max(|R_1|,|R_2|)&
    \text{otherwise},
  \end{cases}
\end{equation}

where $R(x)$ is the position of object $x$ in the ordered set R. $S$ is similar
to the Jaccard index, but, in addition, it penalizes possible differences in
terms of rankings of the elements in the resource vectors.

\begin{figure*}[t]
  \center \includegraphics[width=0.8\textwidth]{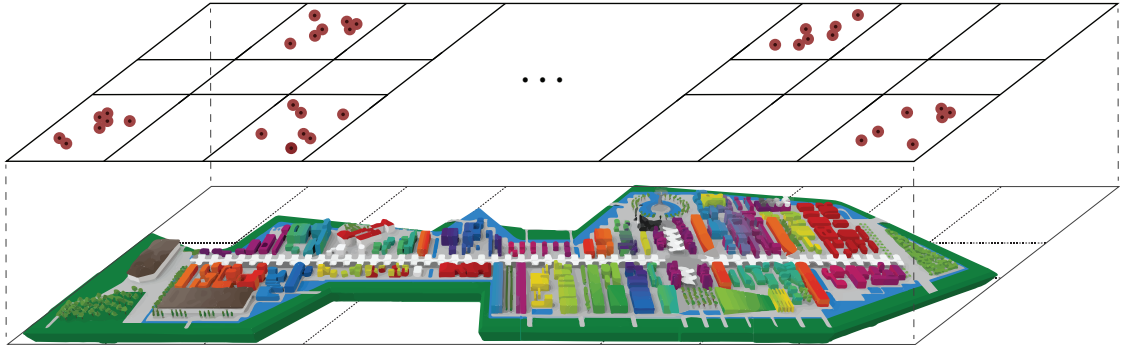}
  \caption{Map of Expo 2015 area with the position of five of the simulated
    communities. Note that the grid in the figure is only an example to show how
    we divided the area for the simulations, but it does not represent the real
    grid used.}
  \label{fig:expo_grid}
\end{figure*}

In addition, for each simulation step, we calculated:

\begin{enumerate}
  \setcounter{enumi}{2}
  \item Number of contents generated by the nodes over time.
  \item Average number of contacts between nodes over time.
\end{enumerate}

These measures are used to characterize the contact traces and the contents used
in the different scenarios. We anticipate that the synthetic and the real traces
that we used show similar properties (e.g., the contact traces used for the
WFD@Expo2015 scenario show values compatible with those used in the conference
scenario), thus supporting the significance of the synthetic trace.

We also calculated all the measures by considering that the interests of nodes
may be limited in time. To do so, we calculated the measures using only the most
recent contents generated in the network and considering only the information
about these contents in the folksonomy graphs. In the simulations, we considered
different ``expiry date'' for the contents, i.e., 1, 2 or 3 hours.

\subsection{Scenario 1 - Big Event: World Food Day @Expo2015}

 As a first dynamic scenario for the evaluation of p-PLIERS, we considered a big
 event attended by a large number of people in a relatively large area. In this
 scenario, accessing the Internet from mobile devices may be problematic and
 thus obtaining useful content from D2D communications would provide an
 important source of information for the users. We considered the World Food Day
 at Expo 2015, organized on October 16, 2015. We assumed that people
 attending the event were able to create tagged contents from their mobile
 devices, and that other attendees might have been interested in obtaining these
 contents through D2D communications.

 \begin{figure*}[t]
  \centering \captionsetup[subfloat]{farskip=2pt,captionskip=3pt}
  \subfloat[\label{fig:expo_user_tweets}]{%
    \includegraphics[width=0.33\textwidth]{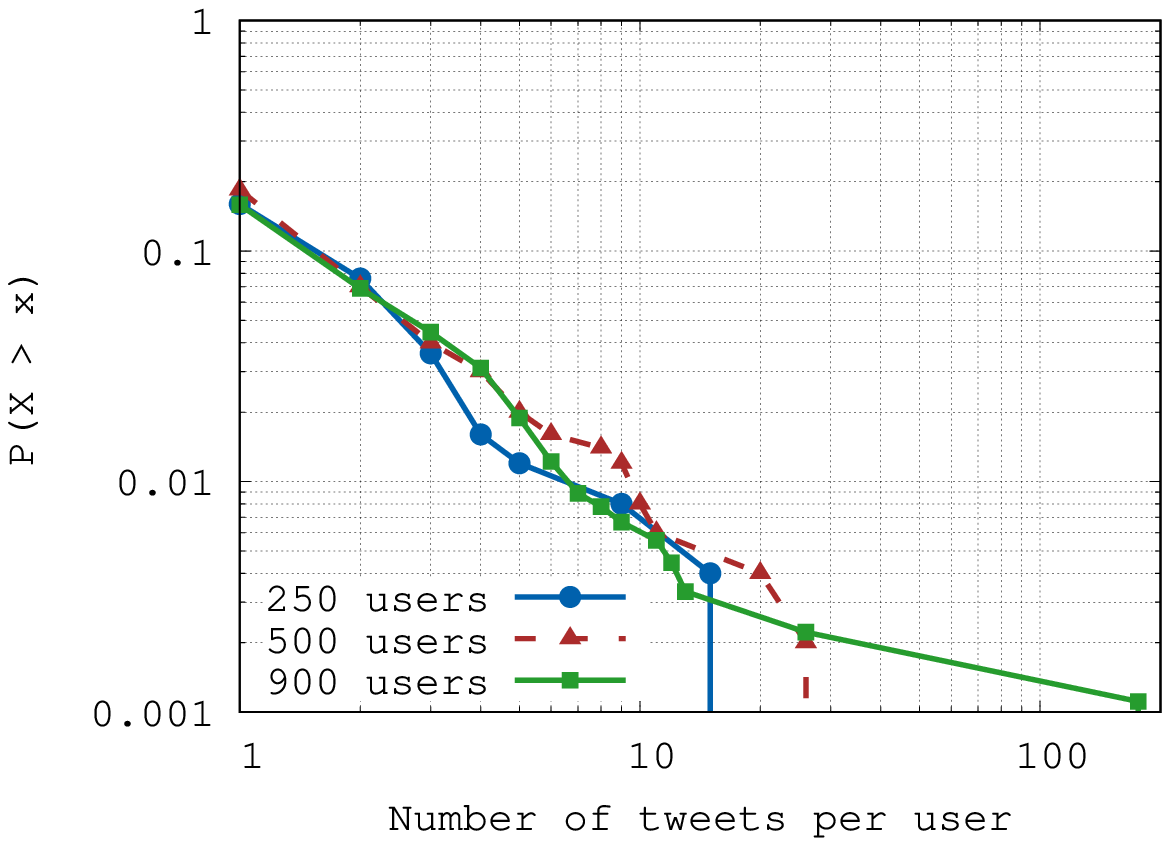} }
  \subfloat[\label{fig:expo_tweet_tags}]{%
    \includegraphics[width=0.33\textwidth]{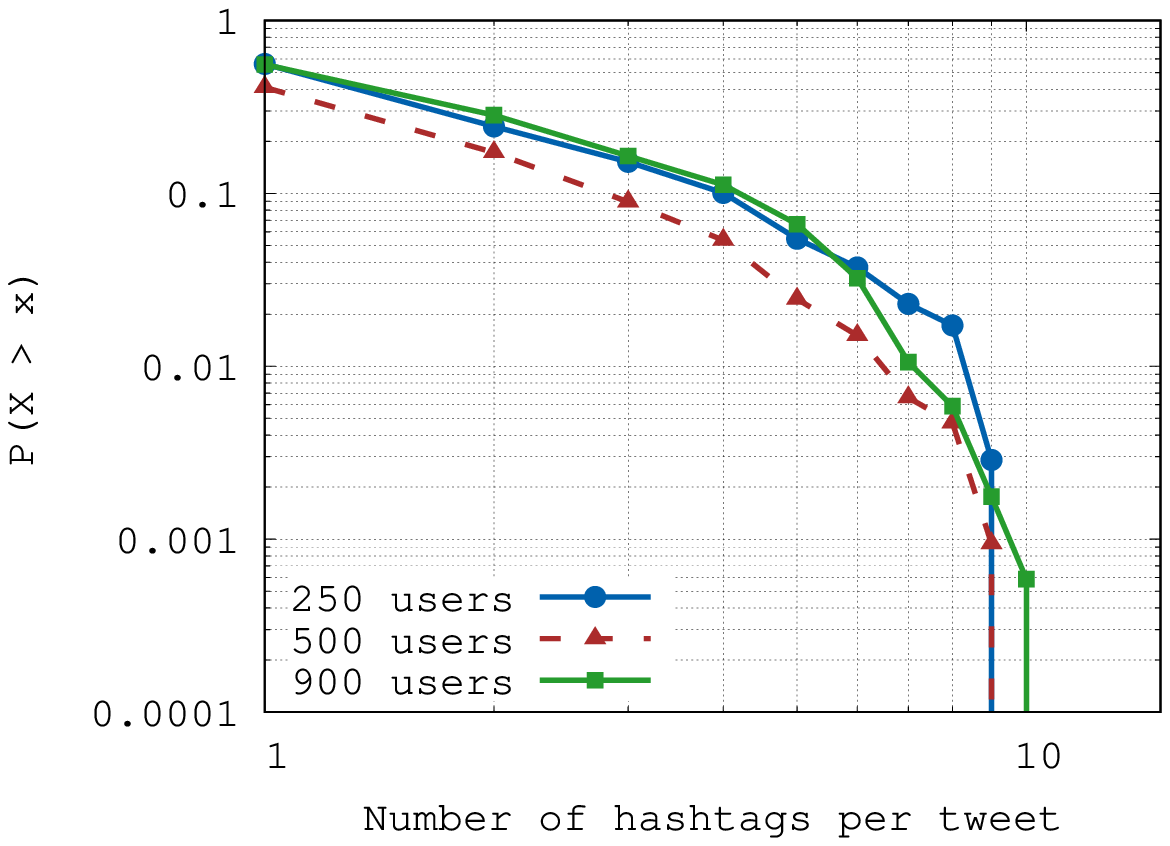} }
  \subfloat[\label{fig:expo_most_used_tags}]{%
    \includegraphics[width=0.33\textwidth]{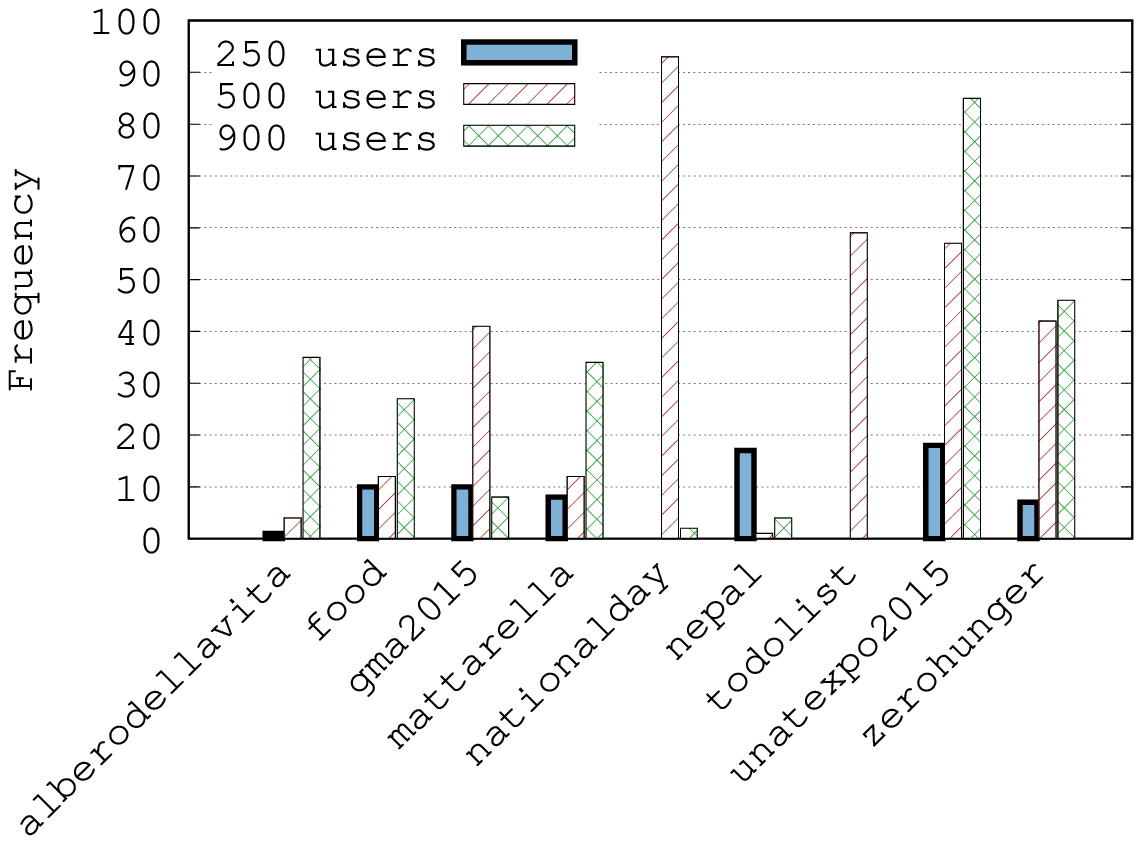} }
  \caption{Descriptive statistics of the Twitter datasets used for the
    WFD@Expo2015 dynamic scenarios. (a) CCDF of the number of tweets per user,
    (b) CCDF of the number of tags per tweet, and (c) Frequency of use of the
    most used tags.}
    \label{fig:helsinki_results}
\end{figure*}

\subsubsection{Mobility Traces}

We simulated a set of nodes moving within an area of 300 x 2,000 meters, which
coincides with the Expo area in Milan. Each node represents a person equipped
with a mobile device. The mobility of nodes is simulated through the HCMM human
mobility model~\cite{boldrini2010hcmm}. The model generates
contacts between nodes according to a Pareto distribution, as observed in real
traces, and considers the presence of ``communities", i.e., each node has a
higher probability to meet nodes within its own community than nodes belonging to
different communities. This fits well with WFD@Expo2015 scenario, as the area
was divided into several pavilions. We can reasonably assume that the mobility
of people inside each pavilion was lower than the mobility of people moving
between pavilions, and that, consequently, the density of intra-community
contacts was higher than that of inter-community ones. We generated mobility
traces through HCMM for 60 communities, approximately the number of pavilions at
Expo (see Figure~\ref{fig:expo_grid} for a graphical representation of a map of
the area of Expo and the communities used in HCMM). We set a probability of
inter-community contacts of 0.1 (this parameter is called ``rewiring"
probability in HCMM)\footnote{We performed the same simulations described in the
  following also with rewiring probabilities 0.2 and 0.3 (thus considering more
  dynamicity in the movements) and this resulted in a general, although slight,
  improvement due to a higher inter-community mobility of the nodes. Since the
  contact traces generated by setting a rewiring probability of 0.1 represent
  the worst case scenario, we present only the results for this
  parameter.}. The speed of nodes ranges between 0.01 m/s (almost steady nodes)
and 1.86 m/s (nodes representing people walking relatively fast).

\begin{figure}[t]
  \center
  \includegraphics[width=0.40\textwidth]{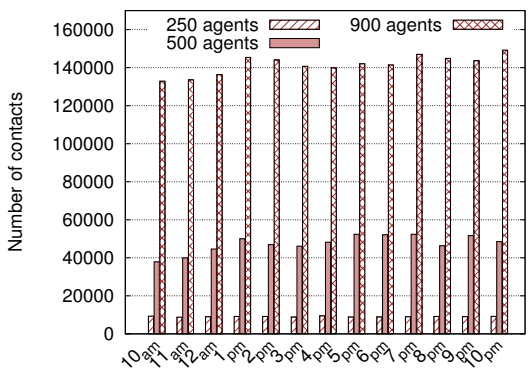}
  \caption{Overall number of contacts between nodes during the simulation for
    the Expo 2015 scenario.}
  \label{fig:expo_contacts_over_time}
\end{figure}

\begin{figure}[t]
  \center
  \includegraphics[width=0.40\textwidth]{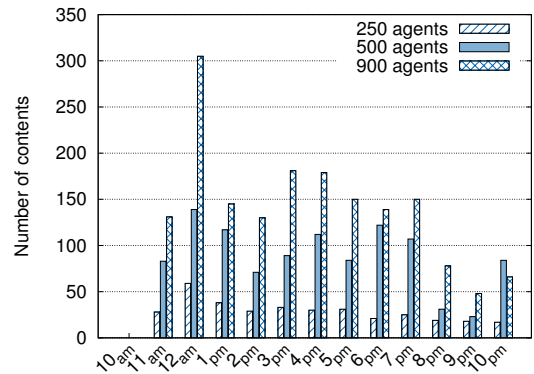}
  \caption{Number of contents generated by nodes during the simulation for the
    Expo 2015 scenario.}
  \label{fig:expo_contents_over_time}
\end{figure}

Using HCMM, we simulated the contacts between nodes for 13 hours, to cover the
timespan of the WFD@Expo2015, from 10am (opening time) till 11pm (closing
time). We considered that each node has a transmission range of 20 meters (the
maximum distance to avoid that nodes in adjacent pavilions are constantly in
contact with each other). When two nodes are in the transmission range of each
other, HCMM generates a contact between them. We think that HCMM mobility model
with these settings can well approximate the real mobility of people during
WFD@Expo2015.

\subsubsection{Content Generation}

To have a realistic representation of multimedia contents generated during the
WFD@Expo2015, we downloaded the tweets generated during the event, using the
Twitter Streaming API. To be sure to obtain only information related to Expo, we
filtered our requests in the same way as done for the dataset collected for the
evaluation in the static scenario presented in
Section~\ref{sec:experiments_static}. With respect to the dataset used for the
static scenario, we downloaded only tweets generated from 10am till 11pm, to
cover the same span of time of the simulated contact traces. The obtained
dataset contains 4,817 tweets generated by 2,660 users, and contains 3,008
unique hashtags.

We associated each node of the simulated contact traces with a Twitter user. In
this way, the tweets of the Twitter user associated with each node represent
contents that the node generated during the simulation.

\begin{figure*}[t]
  \centering \captionsetup[subfloat]{farskip=2pt,captionskip=3pt}
  \subfloat[\label{fig:expo_graph_sim}]{%
    \includegraphics[width=0.33\textwidth]{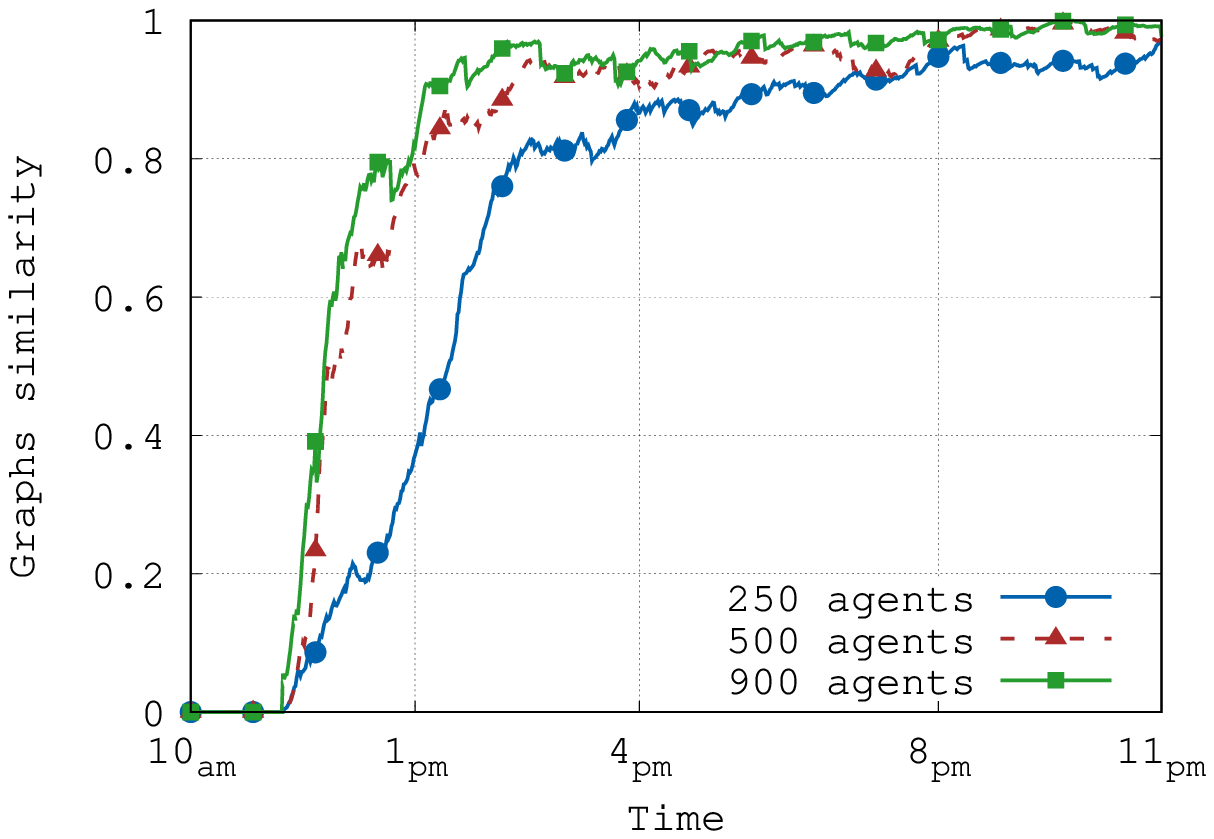} }
  \subfloat[\label{fig:expo_spearman}]{%
    \includegraphics[width=0.33\textwidth]{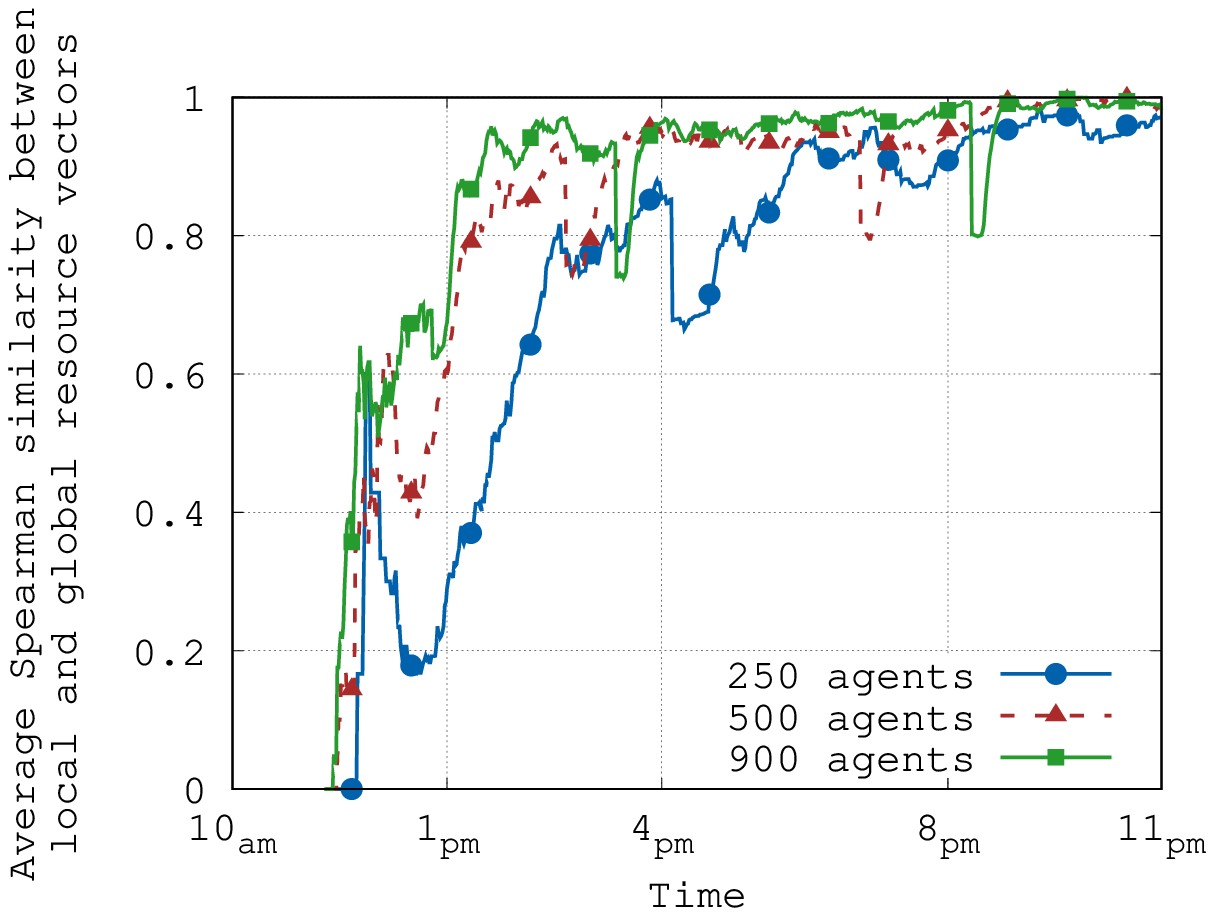} }
  \subfloat[\label{fig:expo_jaccard}]{%
    \includegraphics[width=0.33\textwidth]{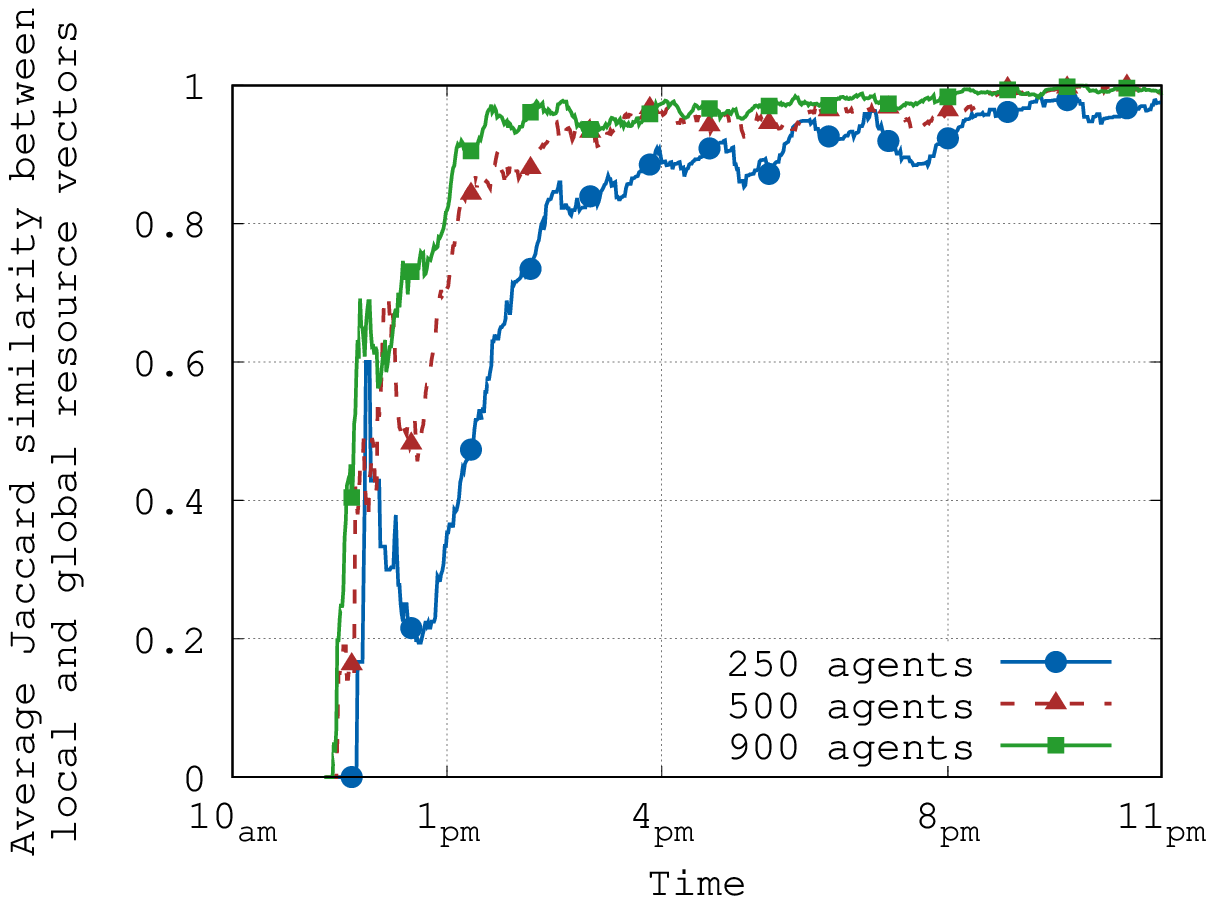} }
    \caption{Results for the WFD@Expo2015 scenario. (a) Average Jaccard
      similarity between local graphs of the agents and the global graph, for
      different number of agents. (b) Average Spearman and (c) Jaccard
      similarity between the PLIERS resource vectors obtained on the local
      graphs of the agents and those obtained from the global graph, for
      different number of agents.}
    \label{fig:expo_results}
\end{figure*}

\subsubsection{Simulation Settings}

We considered different settings for our simulations in order to analyze the
possible impact of different parameters on the results. Specifically, we varied
the number of nodes, simulating 250, 500, and 900 nodes in the considered
area. As the number of users that tweeted during WFD@Expo2015 was higher than
the number of simulated nodes (900 is the maximum number of nodes supported by
HCMM), we sampled, for each setting, the Twitter users with a uniform random
sampling.

In order to give an idea of the dynamics of content generation,
Figures~\ref{fig:expo_user_tweets} and Figure~\ref{fig:expo_tweet_tags} depict
the CCDF of the number of tweets generated by users and the number of hashtags
per tweet respectively, for the samples extracted for the three settings (250,
500, and 900 nodes). The distribution of hashtags per tweet are approximately
the same for the three samples: the majority of tweets have just one or two
hashtags associated with them, whereas just few tweets are linked to more
hashtags. Similarly, the number of users who generated a high number of tweets
is low, when the majority of them has generated just one or two tweets.

Figure~\ref{fig:expo_most_used_tags} depicts the frequency of use for the 5 most
used tags for each sample. As expected, the most used tags are related to topics
about the Expo2015 exhibition, the WFD and some special guests (e.g.,
``mattarella'' refers to Sergio Mattarella, who is the actual President of Italy
and was already in charge during the Expo in 2015, and he was invited as a
special speaker for the WFD event). Figure~\ref{fig:expo_contacts_over_time}
depicts the total number of contacts between nodes for each hour of simulation.
It is worth noting that, for each setting, the number of contacts between nodes is
roughly constant for the entire simulation time, but it greatly differs from one
setting to another, providing thus significantly different cases in terms of
opportunities of contact between nodes in the simulations.

Similarly, Figure~\ref{fig:expo_contents_over_time} depicts the number of
contents (tweets) generated by nodes for each hour of simulation. The
distribution is similar for the three settings: during the lunch break (from
12am till 1pm) there is an increment in the generation of content, then it
remain stable until the evening hours and decreases closer to the exposition
closing time. It is worth noting that no tweets containing the selected hashtags
have been generated between 10am and 11am, probably due to the low number of
visitors in the very first part of the event.

\subsubsection{Results}

\begin{figure*}[t]
  \centering \captionsetup[subfloat]{farskip=2pt,captionskip=3pt}
  \subfloat[\label{fig:kdd_user_tweets}]{%
    \includegraphics[width=0.33\textwidth]{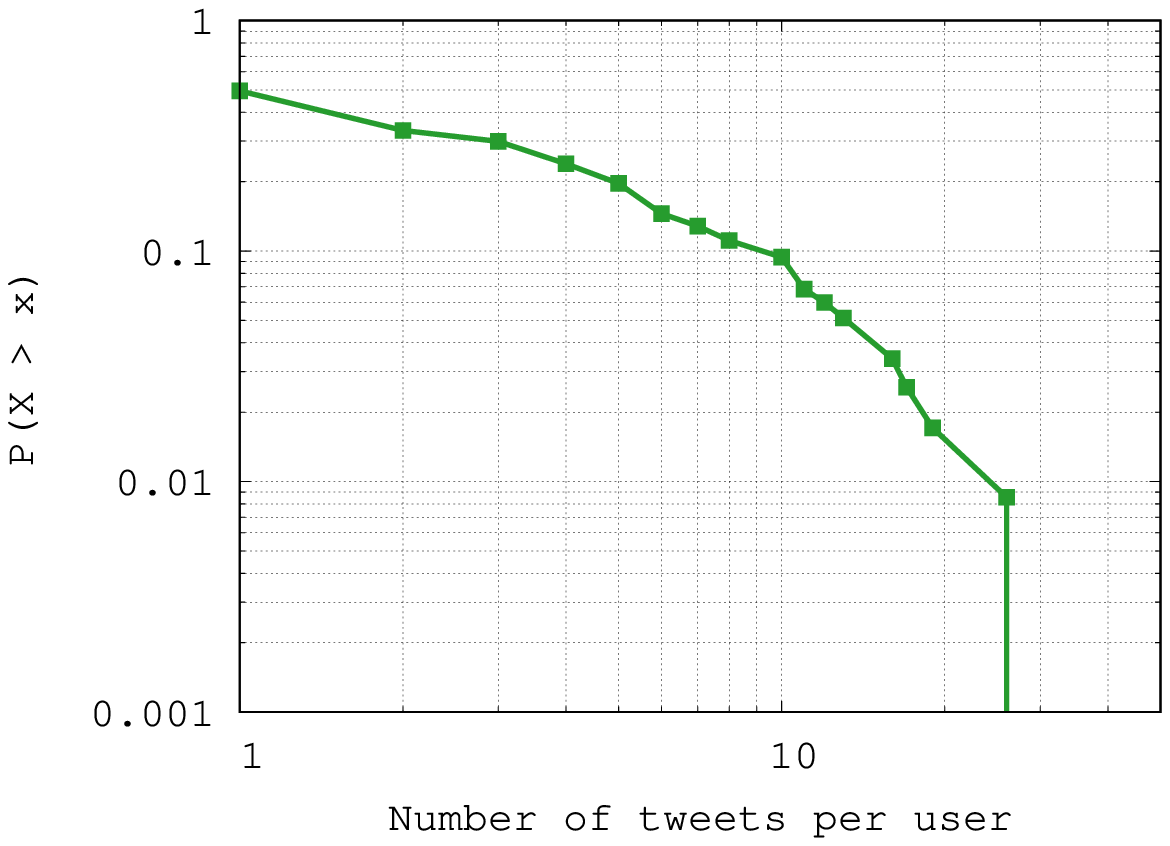} }
  \subfloat[\label{fig:kdd_tweet_tags}]{%
    \includegraphics[width=0.33\textwidth]{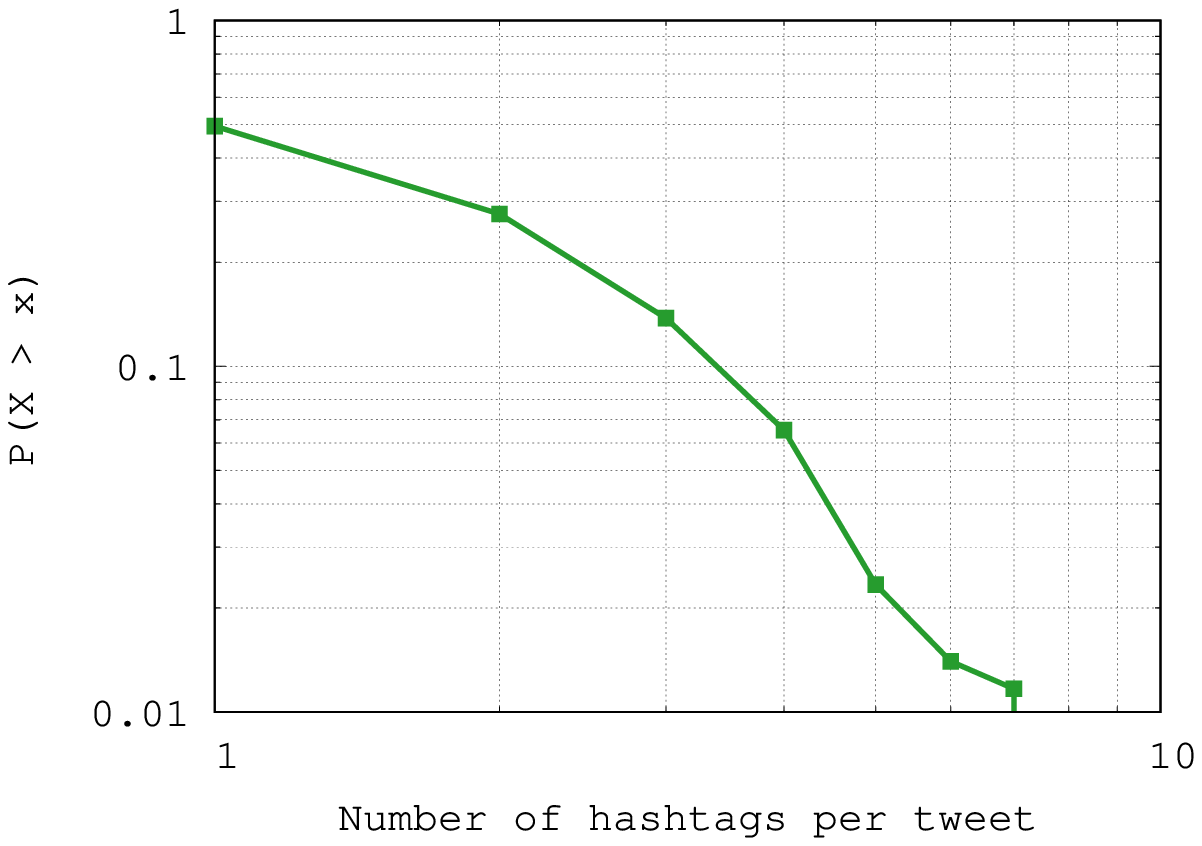} }
  \subfloat[\label{fig:kdd_most_used_tags}]{%
    \includegraphics[width=0.33\textwidth]{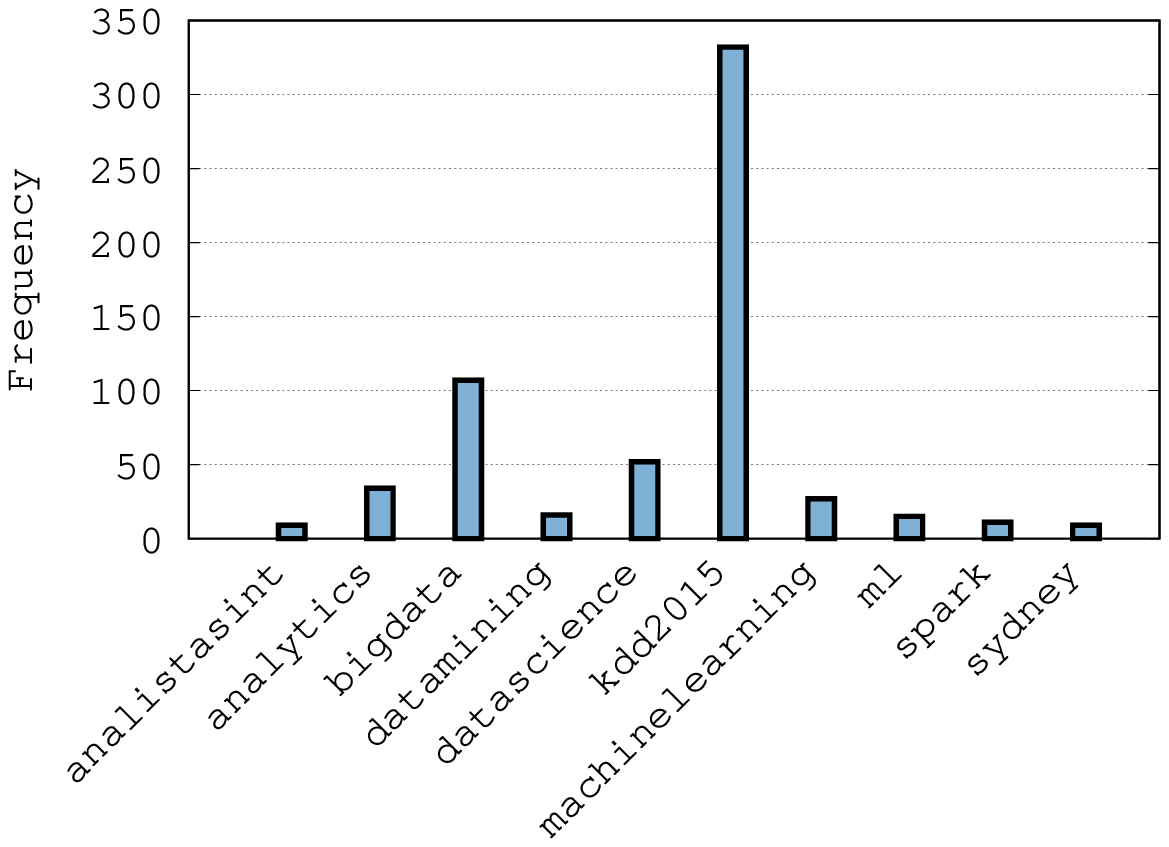} }
  \caption{Descriptive statistics of the Twitter datasets used for the
    Conference scenario: (a) CCDF of the number of tweets per user, (b) CCDF of
    the number of tags per tweet, and (c) Frequency of use of the most used
    tags.}
    \label{fig:kdd_stats}
\end{figure*}

Figure~\ref{fig:expo_graph_sim} depicts the average Jaccard index between the
local graphs of the agents and the global graph, for the different numbers of
simulated agents. From the figure, we can note that, after a certain time, the
similarity between the local graphs and the global graph, on average, reaches a
very high level for all the different settings considered. After approximately
two hours of simulated time, the similarity reaches $\sim$80\% for the cases
with 900 and 500 agents, with slight variations, whereas $\sim$4 hours are
needed to reach the same level of similarity for the case with 250 agents. After
that, the similarity remains quite stable until the end of the simulation for
all the settings. This means that even with a small number of nodes (250) that
generate items in a relatively large area (the whole area of Expo), the
opportunities of contact are enough to have a good approximation of the
knowledge about items and tags generated in the network, even after few
hours. Interestingly enough, the differences between 250 and 900 agents does not
have a substantial impact on the curve.

The similarity between the results obtained by PLIERS on local and global graphs
are depicted in Figure~\ref{fig:expo_spearman}
and~\ref{fig:expo_jaccard}. Specifically, the former figure depicts the average
similarity $S$ (derived from Spearman's Footrule) for the recommendation vectors
of the two cases, while the latter figure depicts the average Jaccard index on
the same recommendation vectors. From the figures, it is worth noting that, on
average, the differences between the results obtained from local and global
graphs are quite small. In addition, the figures clearly indicate that the
results obtained on local graphs converge to those on the global graph. The
presence of negative peaks in Figure~\ref{fig:expo_spearman} that are not
visible in Figure~\ref{fig:expo_jaccard} highlights the higher sensitivity of
the similarity measure $S$ compared to the Jaccard index. These negative peaks
indicate a slight decrease in the performances of PLIERS, which are nonetheless
in the order of $\sim$15\% maximum, and the curve remains always above 0.6 for
all the settings, at least after 4 hours of simulated time.

Figure~\ref{fig:expo_pruning} depicts the average similarity between the local
graphs of the agents and the global graph, where only information generated not
more than 1, 2 and 3 hours (of simulated time) before the calculations is
respectively considered. Note that the figure is related to the simulation with
900 agents. The differences in terms of average similarity of the curves related
to the simulations for 3 and 2 hours are similar to the results obtained without
temporal limitations. This tells us that, nodes can make accurate
recommendations even if they maintain limited information about past history and
purge the older information from their memory. The case where only information
generated in the last hour is considered, the similarity is slightly lower, but
it is still around 75\%, which could be enough for some applications.

\begin{figure}[t]
  \center \includegraphics[width=0.4\textwidth]{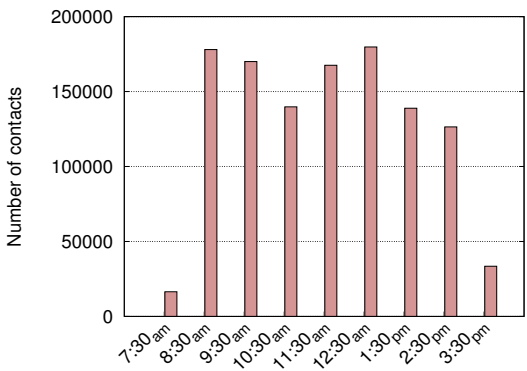}
  \caption{Overall number of contacts between nodes during the simulation for
    the conference scenario.}
  \label{fig:kdd_contacts_over_time}
\end{figure}

\begin{figure}[t]
  \center \includegraphics[width=0.4\textwidth]{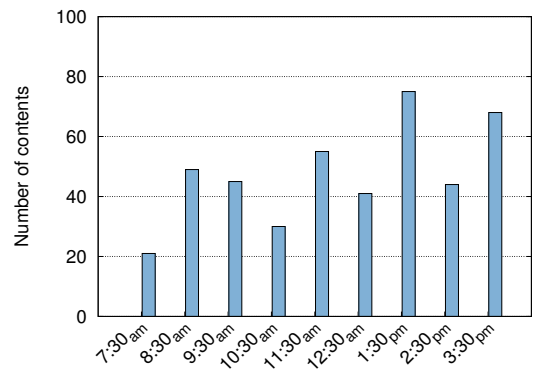}
  \caption{Number of contents generated by nodes during the simulation for the
    conference scenario.}
  \label{fig:kdd_contents_over_time}
\end{figure}

\subsection{Scenario 2 - Conference: KDD 2015}

As a second dynamic scenario, we considered a school campus during a conference
event, where people stay most of the time within rooms, but they regularly
gather at breaks (e.g., coffee or lunch breaks). This is a typical scenario
where pervasive and mobile communications could improve content delivery
services and user experience in general. We assume that the contents generated
during the conference can be shared by nodes through wireless communication, and
we assess the potential impact of PLIERS in the accuracy of recommendations
using only local information obtained by mobile nodes from direct interactions
with other nodes, similarly to the previous scenario.

\subsubsection{Mobility Traces}

For this scenario, we used real contact traces representing the physical
interactions of a group of students, professors, ans staff of an American high
school during a typical school day~\cite{salathe2010high}. These traces were
obtained by distributing 789 wireless sensor network motes to all members of the
school and asking them to carry these motes with them for the entire duration of
a school day.

\subsubsection{Content Generation}

As we do not know the exact location of the American school where the contact
traces have been collected, we tried to find an event from which to collect
tagged online contents and where people were moving in a similar way to how
students move within the area of a high school in the US. As American high
schools are not organized into classes as in the EU, but rather around ``study
tracks'' and students are free to decide which lectures to attend, we think that
their movements can be assimilated to those of people attending a large
conference. For this reason, we downloaded the tweets generated during a large
computer science conference, namely the 21st ACM SIGKDD Conference on Knowledge
Discovery and Data Mining (KDD), organized in Sydney from 10 to 13 August 2015.

To download tweets generated during KDD by people who were physically attending
the conference, we used Twitter REST API with a series of filters, as described
in the following. We first downloaded all the tweets generated by @kdd\_news,
the official Twitter account of the conference. Then, we downloaded all tweets
created by people ``followed'' by this account or ``following'' it and
containing at least one of the set of hashtags generated (in total) by
@kdd\_news. Furthermore, we downloaded additional tweets by performing a second
round of download considering the same set of users (followers and followees of
@kdd\_news) considering the set of hashtags of all the previously downloaded
tweets. In this way, we are not limiting the download to tweets containing only
the hashtags created by the official KDD account, but also those created by
other users that co-occur with the former hashtags. This allowed us to collect
also tweets that are semantically related with the conference. From the set of
downloaded tweets, we kept only those generated during the days of the
conference (10-13 August) and which are within the temporal span between 7.30am
and 4.30pm Sydney time, coinciding with the intersection between the time window
of the conference and that of the contact traces.

The total number of KDD tweets that we downloaded is 428 (distributed over the
three conference days). These tweets have been created by 117 users, and contain
a total 256 unique hashtags.

\begin{figure*}[t]
  \centering \captionsetup[subfloat]{farskip=2pt,captionskip=3pt}
  \subfloat[\label{fig:kdd_graph_sim}]{%
    \includegraphics[width=0.33\textwidth]{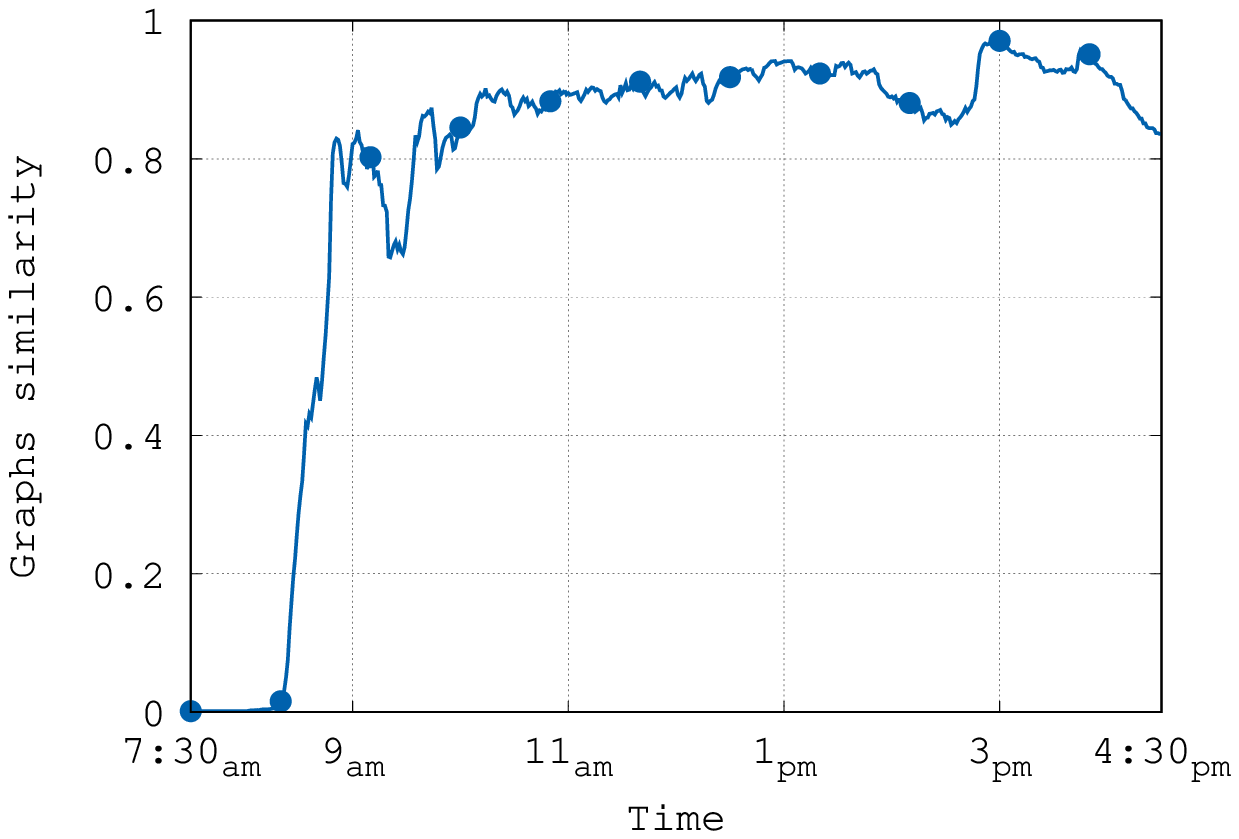} }
  \subfloat[\label{fig:kdd_spearman}]{%
    \includegraphics[width=0.33\textwidth]{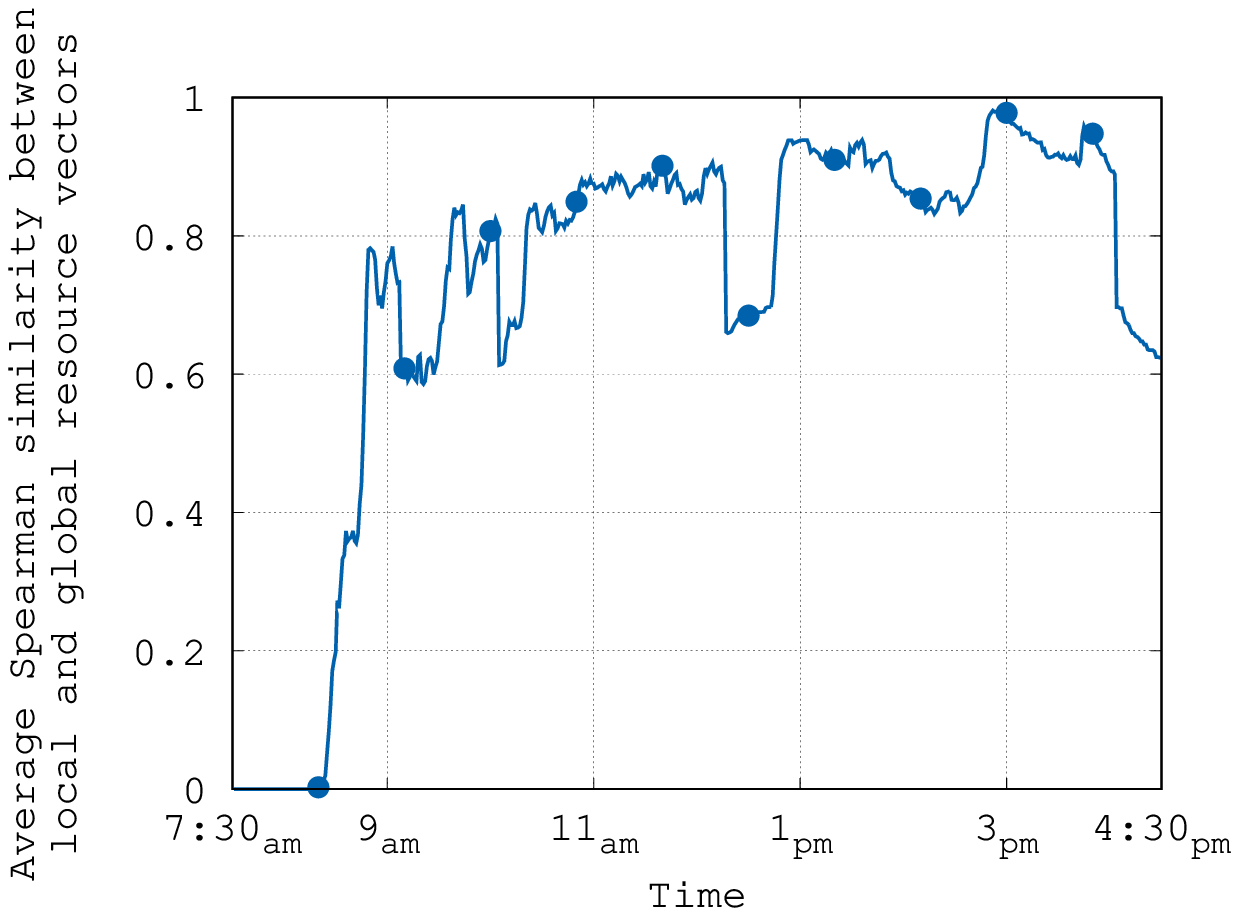} }
  \subfloat[\label{fig:kdd_jaccard}]{%
    \includegraphics[width=0.33\textwidth]{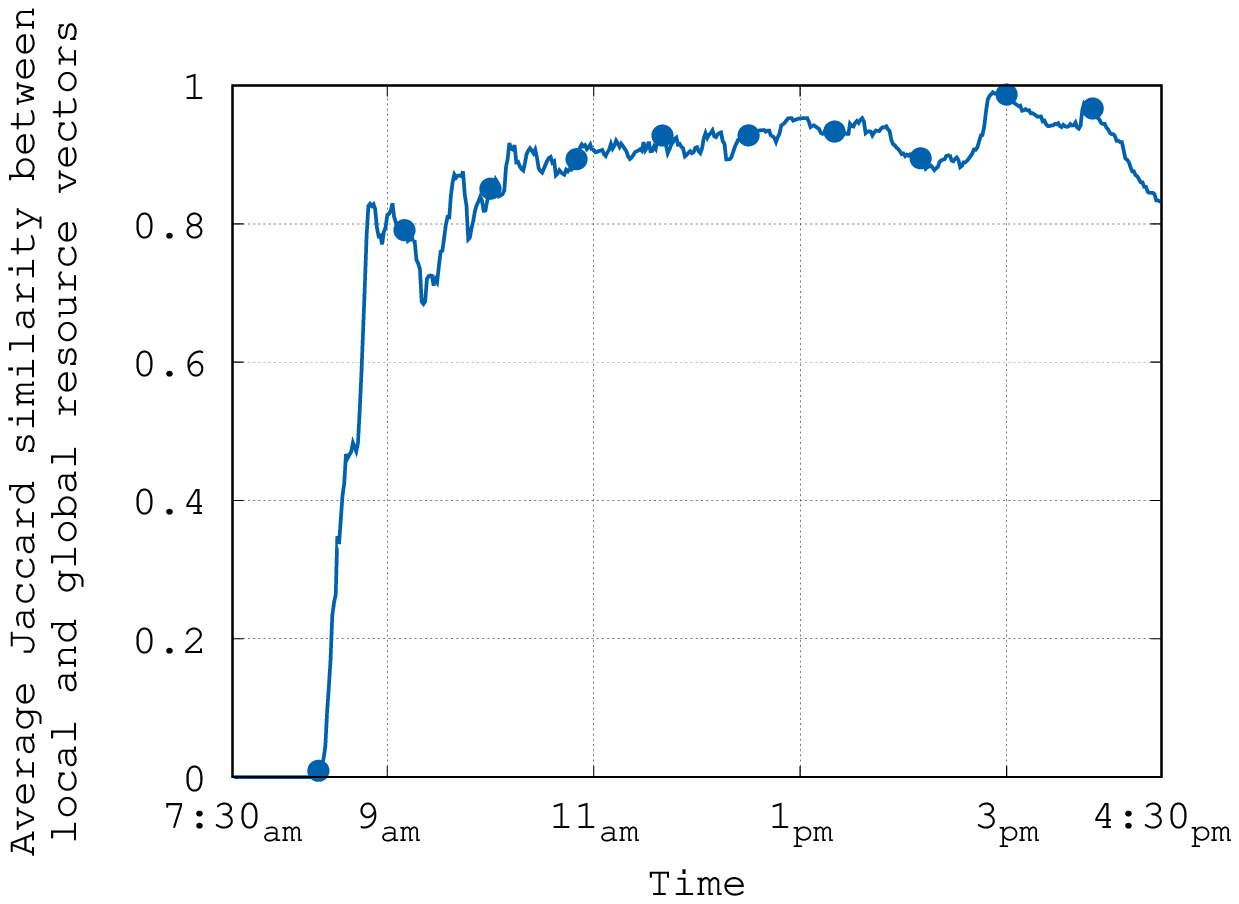} }
    \caption{Results for the scenario of the KDD conference. (a) Average Jaccard
      similarity between local graphs of the agents and the global graph, for
      different number of agents. (b) Average Spearman and (c) Jaccard
      similarity between the PLIERS resource vectors obtained on the local
      graphs of the agents and those obtained from the global graph, for
      different number of agents.}
    \label{fig:kdd_results}
\end{figure*}

\subsubsection{Simulation Settings}

Since the total number of Twitter users in our dataset is lower than the number
of nodes in the contact traces, we decided to assign each Twitter user to a
random node in the simulator. The nodes not associated with a Twitter user do
not create any contents, but they are still part of the simulation and they can
diffuse information within the mobile network. In addition, to maintain a
sufficiently high number of generated contents, we considered that the tweets
were generated on a single day, and we just considered the creation time of each
tweet, and not its creation date.

Figure~\ref{fig:kdd_user_tweets} and Figure~\ref{fig:kdd_tweet_tags} depict
respectively the CCDF of the number of tweets generated per node and that of the
number of tags per tweet considered in the simulation. The tags appearing with
highest frequency during the simulation are depicted in
Figure~\ref{fig:kdd_most_used_tags}. The number of contacts and the number of
contents generated hourly during the simulation are depicted in
Figure~\ref{fig:kdd_contacts_over_time} and
Figure~\ref{fig:kdd_contents_over_time} respectively.
Figure~\ref{fig:kdd_contacts_over_time}. shows a number of contacts of the same
order of magnitude as WFD@Expo15 for a comparable number of users.

\subsubsection{Results}

Figure~\ref{fig:kdd_graph_sim} depicts the average similarity between local and
global graphs for the simulation. The high value of similarity indicates that
nodes are able, on average, to have a complete view of the contents around them
even after a few hours of simulated time. In addition, also the similarity
between the recommendations of PLIERS obtained by the nodes and the optimal
recommendations they would have got using the global knowledge is very high both
for the $S$ measure (see Figure~\ref{fig:kdd_spearman}) and for the Jaccard
similarity index (Figure~\ref{fig:kdd_jaccard}).

The results obtained by limiting information lifetime at 1, 2, and 3 hours for
this scenario are reported in Figure~\ref{fig:kdd_pruning}. In this case, the
results obtained for the most restrictive assumptions, considering only
information generated 1 and 2 hours respectively before each step of the
simulation, have a high variation and often go beyond a similarity of 0.6. The
similarity in these cases might be too low to obtain meaningful
recommendations. Nevertheless, the results obtained for the threshold of 3 hours
are very similar to those obtained without limiting the information
lifetime. This indicates that, for this kind of scenario, a view of the contents
generated within the last 3 hours could be enough to obtained accurate
recommendations about freshly created information.

\subsection{Scenario 3 - City: Helsinki}

As a last scenario to test our solution, we chose the urban environment of the
city center of Helsinki, a medium sized European city. Our choice was motivated
by the need to test PLIERS on a larger scale than the previous scenarios. We
extracted the contact traces of a typical working day in Helsinki using a
realistic human mobility model highly customized on the considered area. Then,
we downloaded the tweets generated within the same geographic area and we used
them as content generated by nodes during the simulation.

\subsubsection{Mobility Traces}

\begin{figure}[t]
  \center \includegraphics[width=0.48\textwidth, trim={2cm 2.3cm 2.3cm
      1cm},clip]{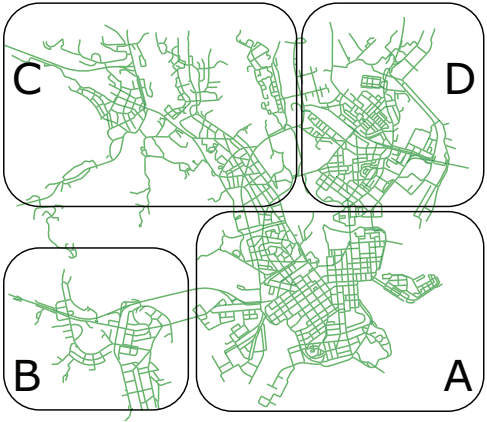}
  \caption{Graphical representation of the four groups of nodes in the mobility
    model of Helsinki which live and work in the same districts of the city.}
  \label{fig:helsinki_abcd}
\end{figure}

\begin{figure}[t]
  \center \includegraphics[width=0.41\textwidth, trim={2cm 1.5cm 2.3cm
      0cm},clip]{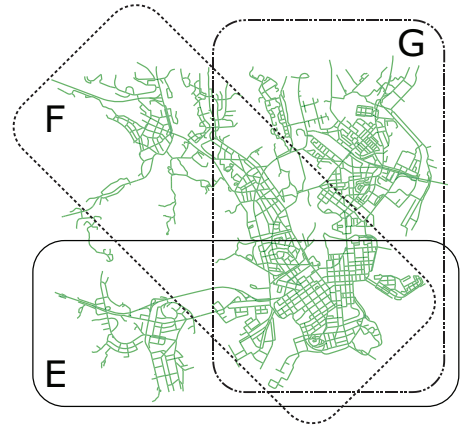}
  \caption{Graphical representation of the additional three groups of nodes in
    the mobility model of Helsinki which live and work in different districts of
    the city.}
  \label{fig:helsinki_efg}
\end{figure}

In order to perform a realistic simulation, we extracted 24 hours of contacts
between 800 nodes from the mobility traces generated by the working day mobility
model for Helsinki~\cite{ekman2008working} implemented in TheONE simulator. The
mobility model uses several highly customized mobility sub-models that define
nodes behavior during different daily activities in the area of interest, such
as staying at home, working, and evening activities with friends. The simulation
area coincides with Helsinki city center, which was divided into four main
districts, as depicted in Figure~\ref{fig:helsinki_abcd}. In these districts,
nodes live and work in the same geographic area forming thus four distinct
groups. In addition, the model considers three other groups of nodes which live
and work in different districts, as depicted in
Figure~\ref{fig:helsinki_efg}. To simulate the movements between home and work,
and between work and possible meeting points for evening activities, the model
defines three additional mobility models: car travel mobility, public
transportation mobility, and walking mobility. For a complete description of
each of the mobility sub-models included in the overall mobility model and to
understand how they are combined together, we refer the reader
to~\cite{ekman2008working}. For the parameters of the models, we used the
default values provided by TheONE simulator. Note that the values were directly
derived for the city of Helsinki.

\subsubsection{Content Generation}

We downloaded the tweets generated in the area of Helsinki city center through
Twitter Streaming API, by filtering the tweets for their location. We
continuously downloaded tweets from May 27, 2016 to June 20, 2016, for a total
of 24,732 tweets, which were generated by 4,477 users and 16,273 unique
hashtags.

\subsubsection{Simulation Settings}

We performed three simulations using the same contact traces obtained by the
Helsinki mobility model and varying the number of contents generated by each
node. To do so, we selected, in the first case, only the tweets generated during
a single working day randomly chosen among those in the collected dataset (May
2, 2016 - which was a typical working day in Helsinki). The number of Twitter
users who generated contents on this day was 586. We randomly assigned these
users to the 800 nodes of the contact traces. The remaining nodes do not
generate contents, but they contribute to the forwarding of knowledge in the
network. For the second simulation, we first selected all the users in the
dataset who tweeted for at least two days (not necessarily consecutive
days). Then, we randomly selected two days of tweeting activity for each user
(filtering out additional tweets for the users who tweeted for more than two
days). As the number of users was higher than 800, we simply randomly assigned
each node in the contact traces to a randomly selected Twitter user. For the
third simulation, we performed the same selection of the previous case, but for
three days of tweeting activity.

\begin{figure*}[t]
  \centering \captionsetup[subfloat]{farskip=2pt,captionskip=3pt}
  \subfloat[\label{fig:helsinki_user_tweets}]{%
    \includegraphics[width=0.33\textwidth]{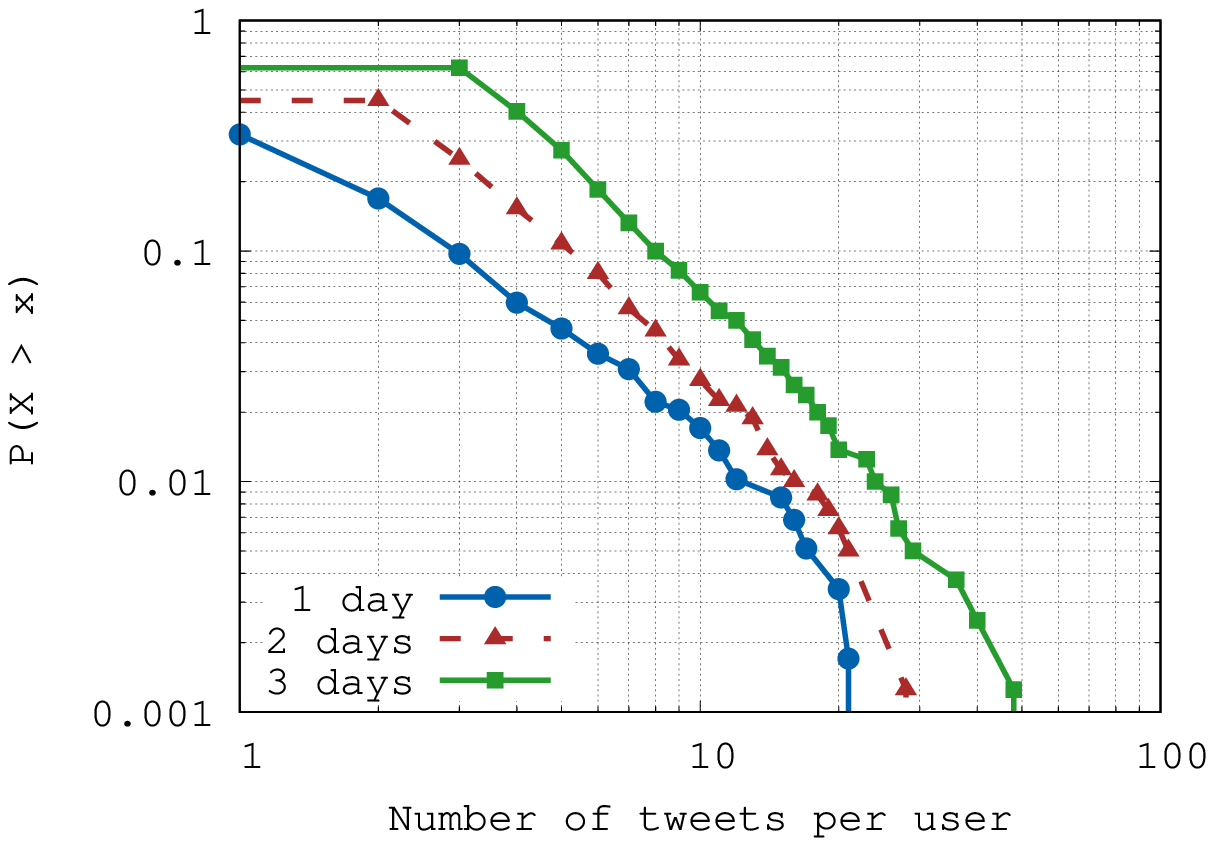} }
  \subfloat[\label{fig:helsinki_tweet_tags}]{%
    \includegraphics[width=0.33\textwidth]{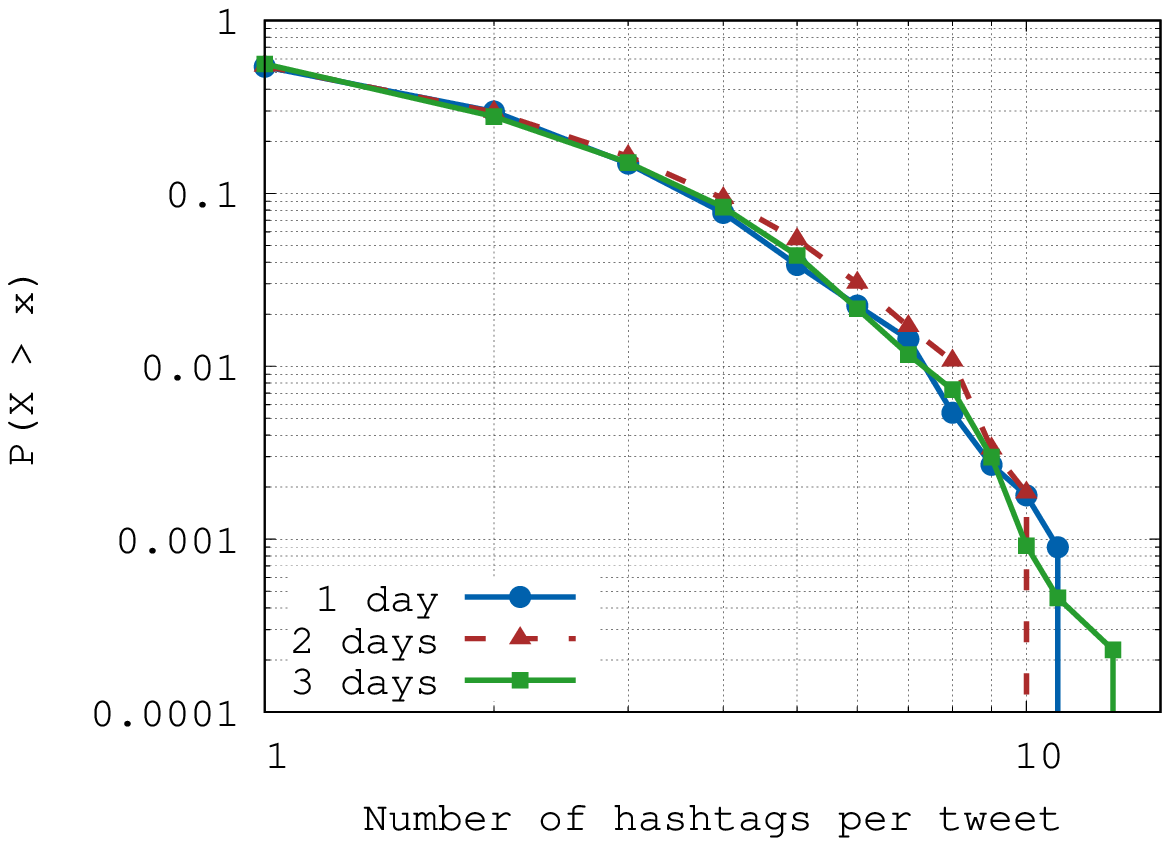} }
  \subfloat[\label{fig:helsinki_most_used_tags}]{%
    \includegraphics[width=0.33\textwidth]{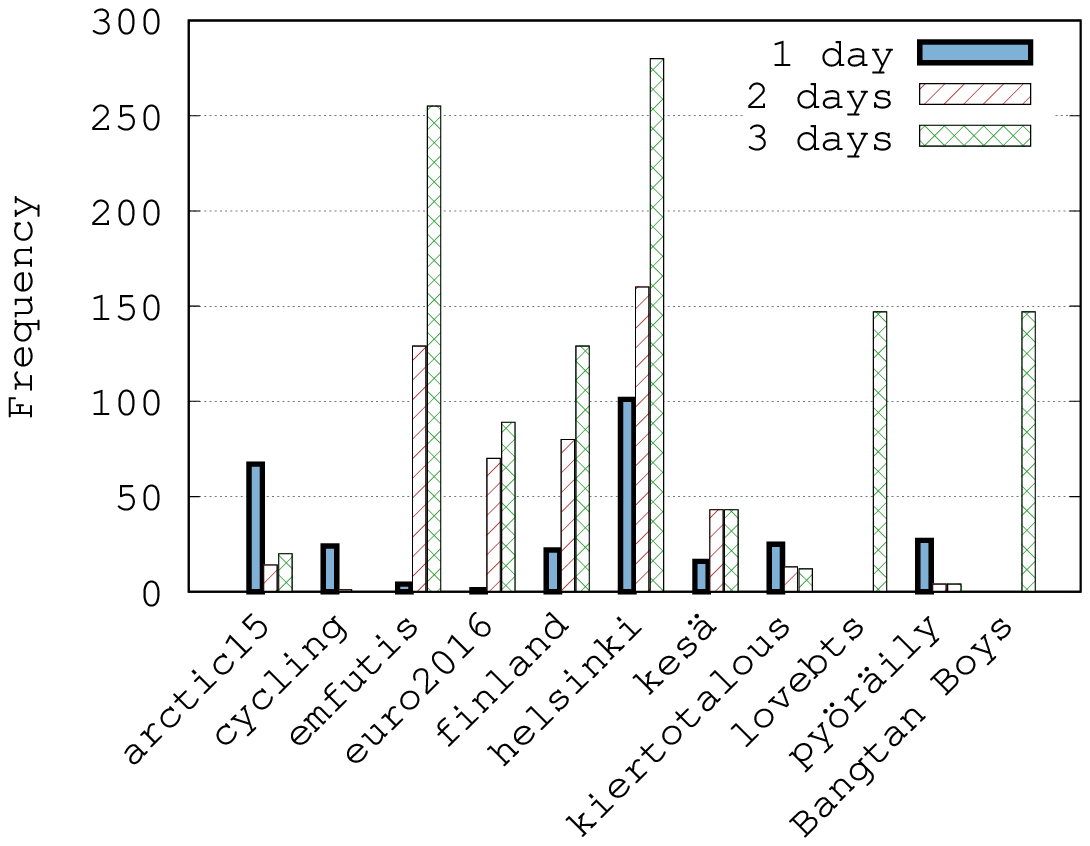} }
  \caption{Descriptive statistics of the Twitter datasets used for the City
    scenario: (a) CCDF of the number of tweets per user, (b) CCDF of the number
    of tags per tweet, and (c) Frequency of use of the most used tags.}
    \label{fig:helsinki_stats}
\end{figure*}

\begin{figure}[t]
  \center
  \includegraphics[width=0.45\textwidth]{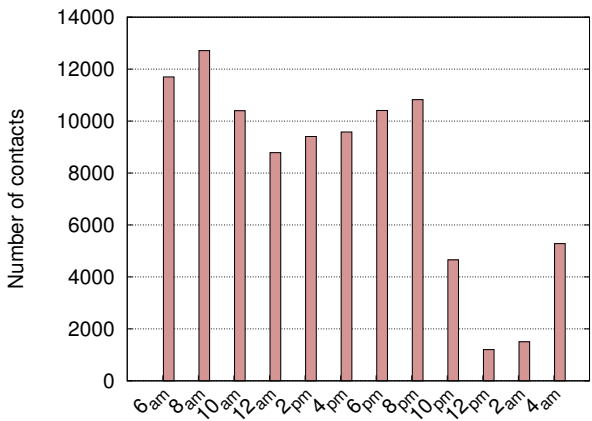}
  \caption{Number of contacts generated by nodes during the simulation.}
  \label{fig:helsinki_contacts_over_time}
\end{figure}

\begin{figure}[t]
  \center
  \includegraphics[width=0.45\textwidth]{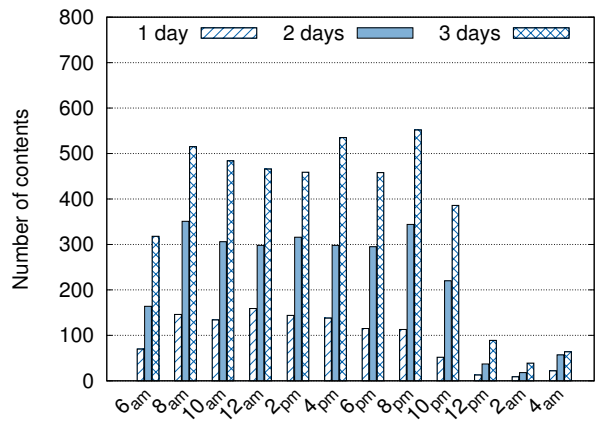}
  \caption{Number of contents between nodes during the simulation.}
  \label{fig:helsinki_contents_over_time}
\end{figure}

 Figure~\ref{fig:helsinki_user_tweets} and Figure~\ref{fig:helsinki_tweet_tags}
 depict the CCDF of the number of tweets generated per node and the CCDF of the
 number of tags per tweet for this scenario, considering the three settings with
 contents generated over 1, 2, and 3 days. The tags with highest frequency for
 the three cases are depicted in Figure~\ref{fig:helsinki_most_used_tags}. The
 number of contacts and the number of contents generated hourly during the
 simulation are depicted in Figure~\ref{fig:helsinki_contacts_over_time} and
 Figure~\ref{fig:helsinki_contents_over_time} respectively.

\subsubsection{Results}

\begin{figure*}[t]
  \centering \captionsetup[subfloat]{farskip=2pt,captionskip=3pt}
  \subfloat[\label{fig:helsinki_graph_sim}]{%
    \includegraphics[width=0.33\textwidth]{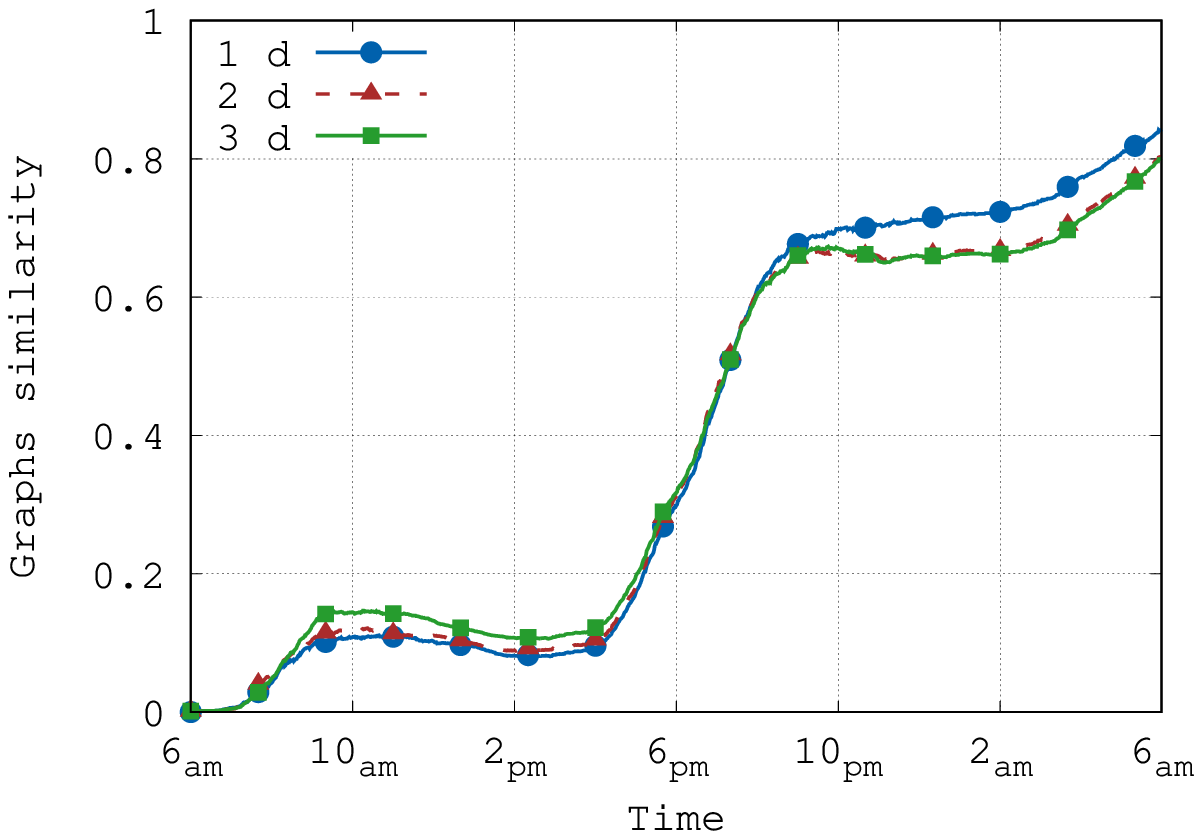} }
  \subfloat[\label{fig:helsinki_spearman}]{%
    \includegraphics[width=0.33\textwidth]{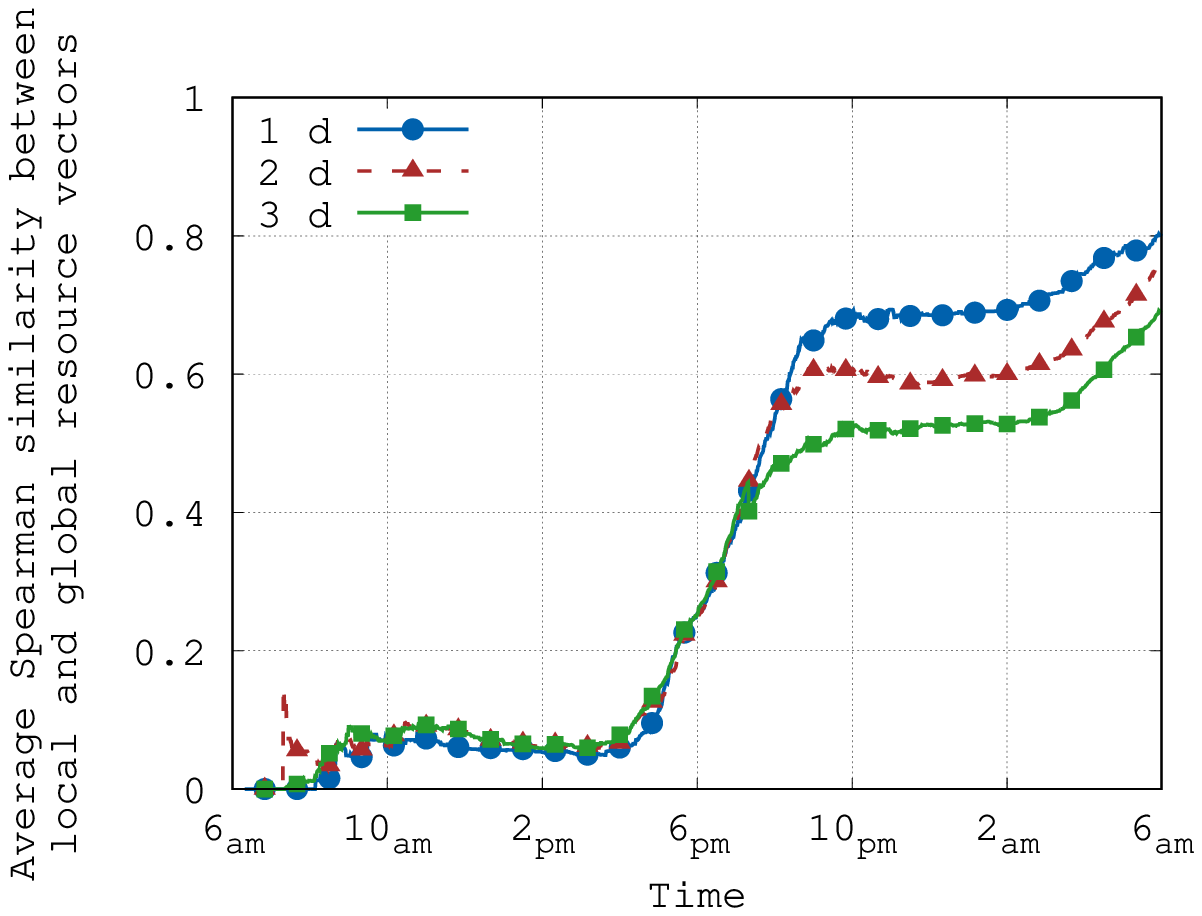} }
  \subfloat[\label{fig:helsinki_jaccard}]{%
    \includegraphics[width=0.33\textwidth]{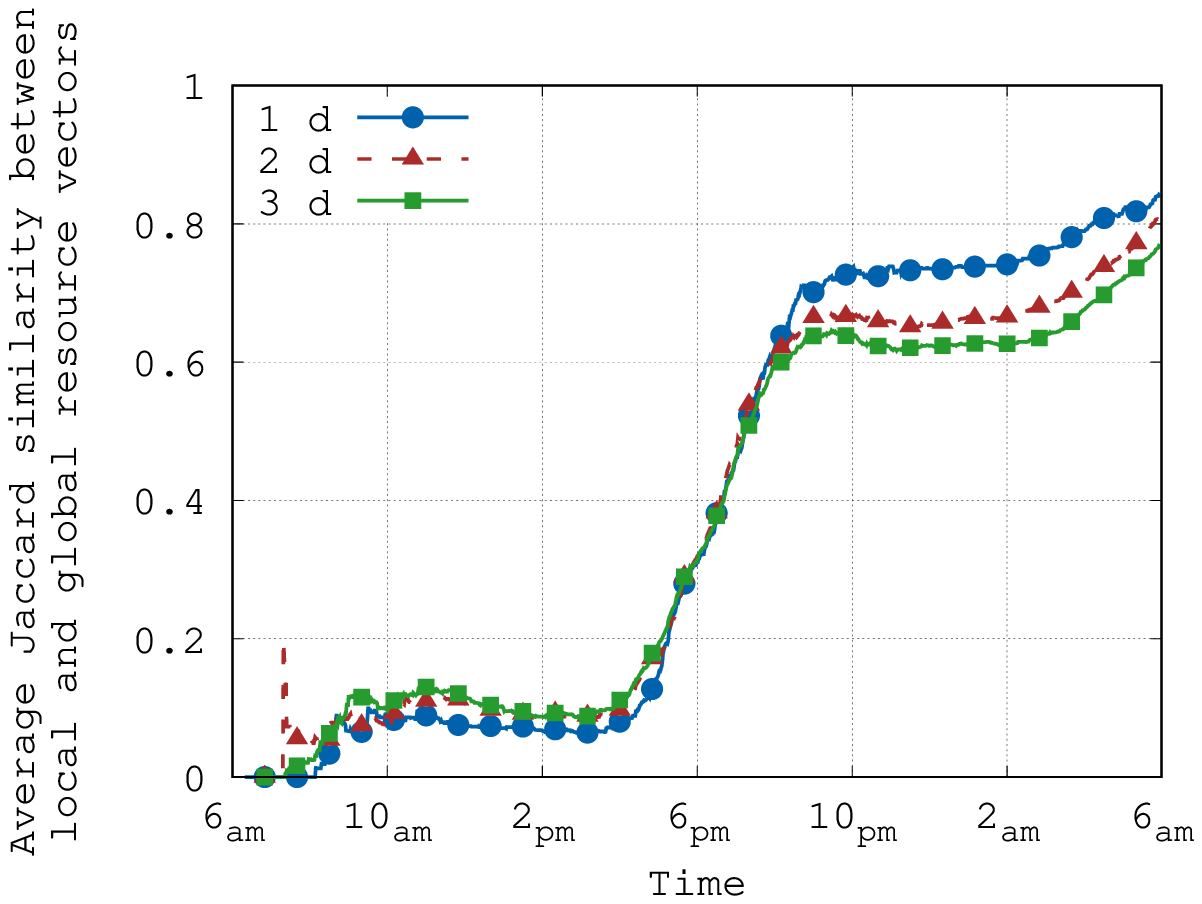}
  }
    \caption{Results for the scenario of the city center of Helsinki. (a)
      Average Jaccard similarity between local graphs of the agents and the
      global graph, for different number of agents. (b) Average Spearman and (c)
      Jaccard similarity between the PLIERS resource vectors obtained on the
      local graphs of the agents and those obtained from the global graph, for
      different number of agents.}
    \label{fig:helsinki_results}
\end{figure*}

Figure~\ref{fig:helsinki_graph_sim} depicts the average similarity between local
and global graphs during the three simulations. In the first half of the day the
similarity is rather low (i.e., less than 20\%). This is due to the fact that
the agents are mostly found at home or work, and then they do not have many
opportunities to meet new nodes from which they could get new information about
contents generated in the network. When the agents stop working approximately at
5:30 p.m., the graphs similarity rapidly grows to over the 80\%. This is because
most of the agents go to the meeting points (e.g., shopping center, restaurants,
pubs, etc\ldots), which allow them to encounter nodes from different communities
and then become aware of the content never seen before.

Figures~\ref{fig:helsinki_spearman} and~\ref{fig:helsinki_jaccard} depict
respectively the average Spearman and Jaccard similarity between the
recommendations of PLIERS obtained by the nodes and the optimal recommendations
they would have got using the global knowledge. It is worth noting that the
curves of the recommendations' similarity accurately reflect the similarity
between the local and global graphs. This proves that the higher the accuracy of
the vision of the local nodes and the greater will be the accuracy of the
recommendations made ​​by PLIERS.

\begin{figure*}[t]
  \centering \captionsetup[subfloat]{farskip=2pt,captionskip=3pt}
  \subfloat[\label{fig:expo_pruning}]{%
    \includegraphics[width=0.33\textwidth]{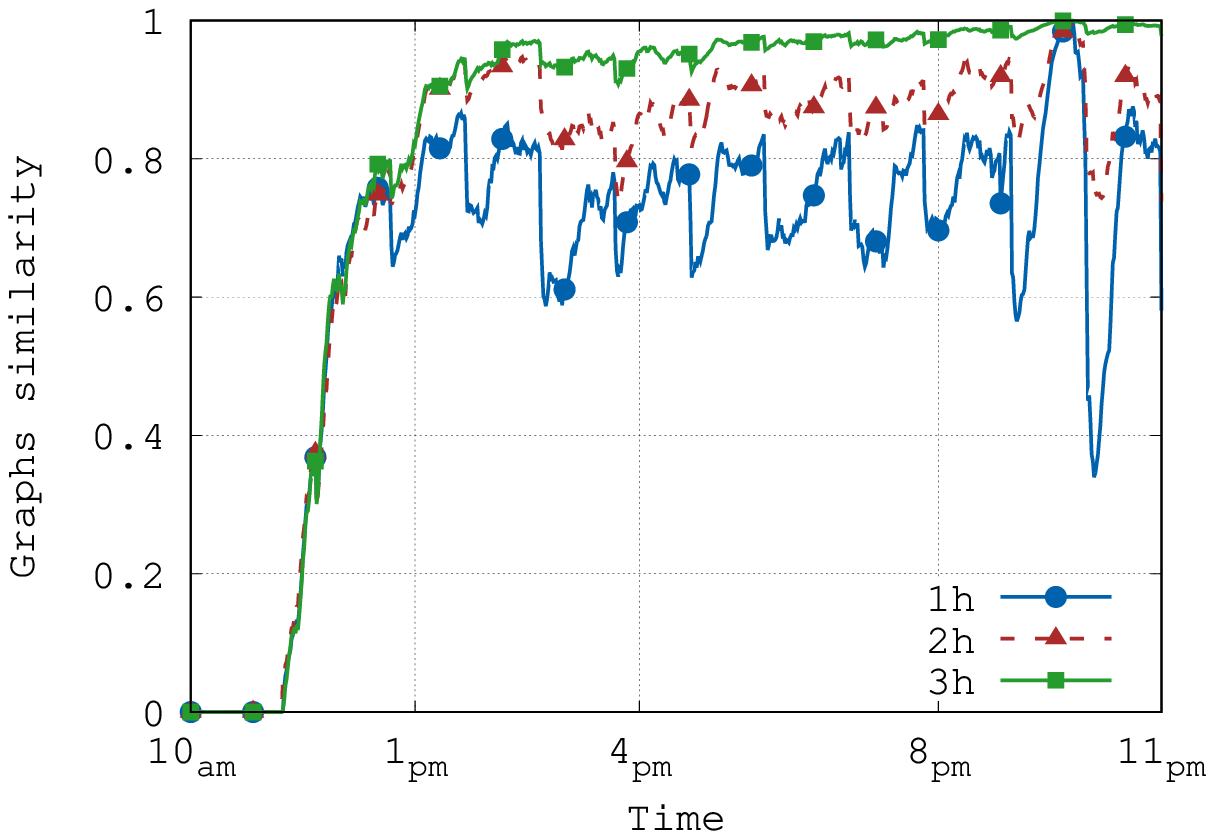} }
  \subfloat[\label{fig:kdd_pruning}]{%
    \includegraphics[width=0.33\textwidth]{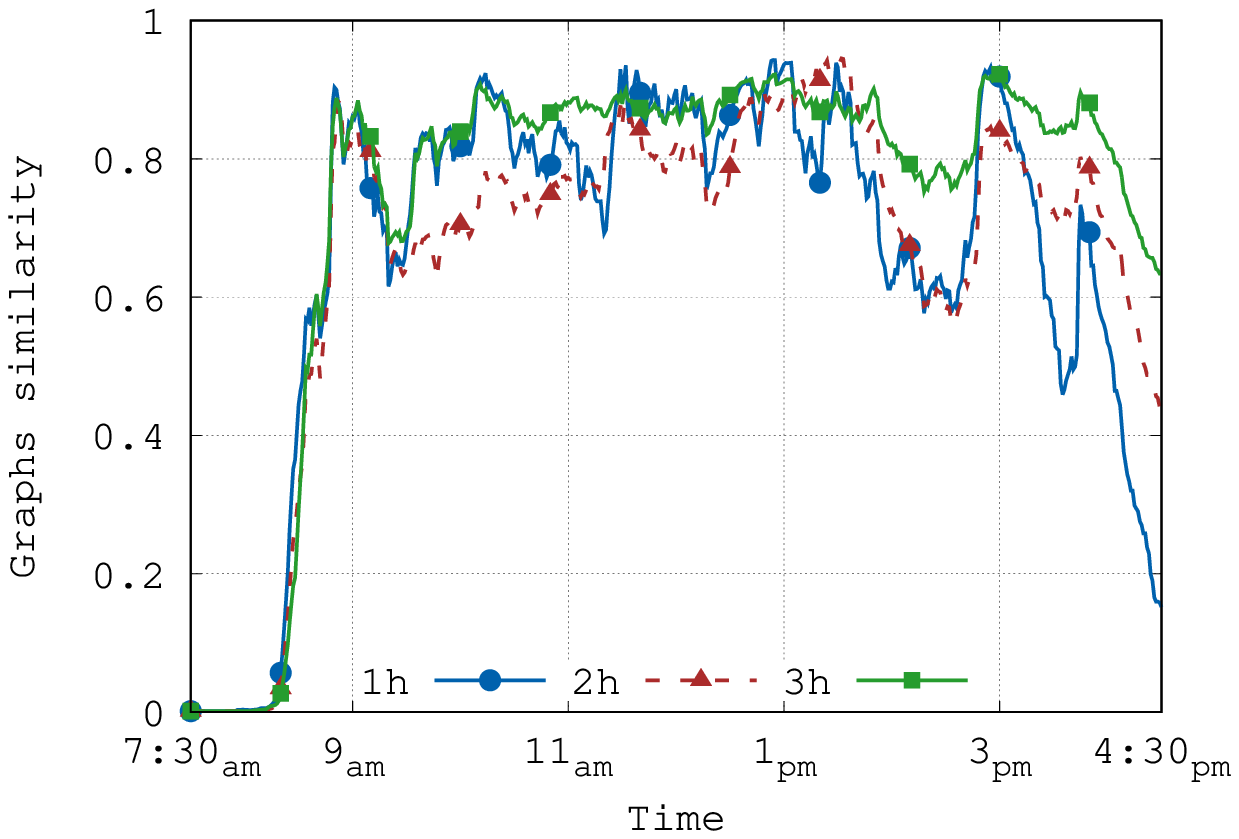} }
  \subfloat[\label{fig:helsinki_pruning}]{%
    \includegraphics[width=0.33\textwidth]{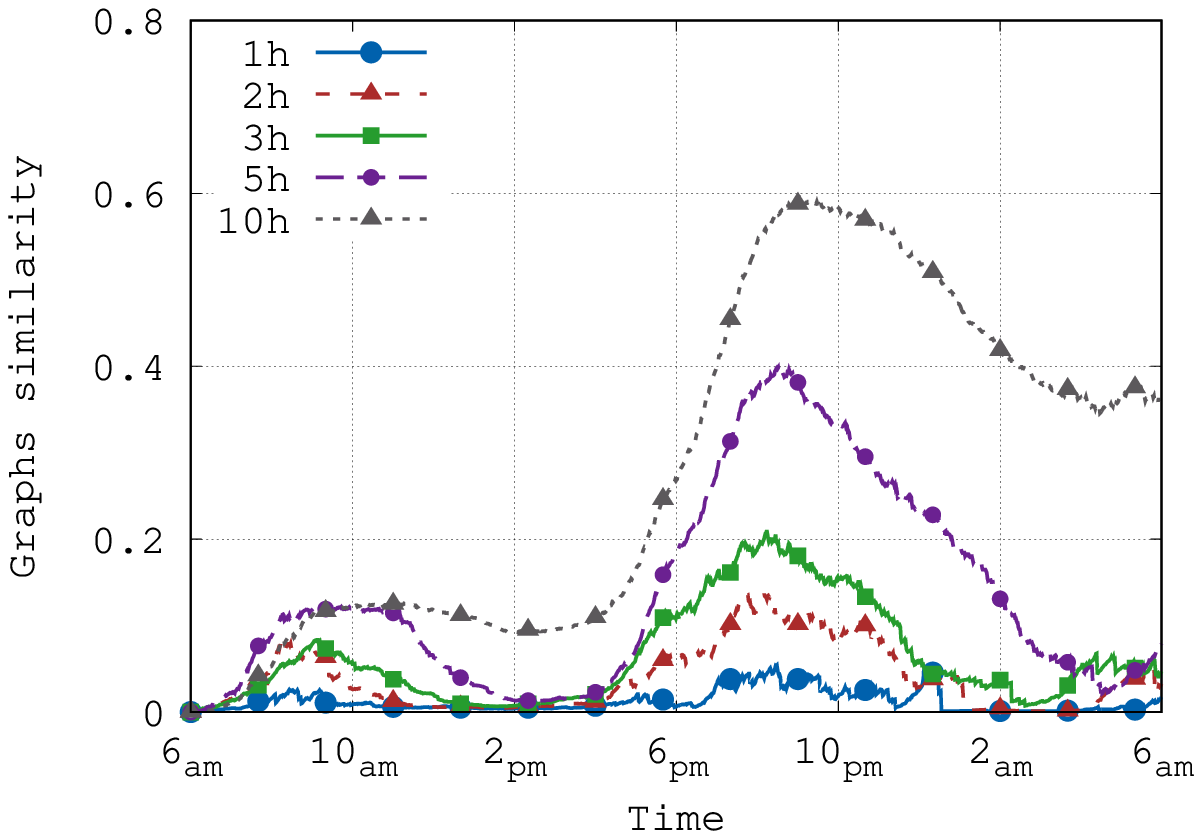} }
    \caption{Average Jaccard similarity between local and global graphs,
      limiting the knowledge to different time windows in the past for the (a)
      Expo2015, (b) Conference, and (c) Helsinki scenarios.}
    \label{fig:pruning}
\end{figure*}

The results obtained by limiting information lifetime at different hours for
this scenario are reported in Figure~\ref{fig:helsinki_pruning}. In this case,
the thresholds used in the other scenarios to limit the knowledge graphs are too
restrictive, and the similarity between the local and global graphs remains
beneath of the 20\%. For this reason, we used two higher threshold values: we
considered only the information generated 5 and 10 hours before each step of the
simulation. Considering the information generated during approximately half of
the simulation time (i.e., 10 hours), the similarity between the local and
global graphs considerably increases, and reaches 60\%. This result suggests
that, for a realistic urban scenario, at least a half day of knowledge about the
contents generated by nodes is necessary to obtain sufficiently accurate
results.

\subsection{Discussion}

The results for the three dynamic scenarios show that p-PLIERS is able to adapt
to several types of realistic pervasive social networks and always provides
accurate content recommendations.

We performed an additional analysis to verify that the results are not trivially
related to the number of contacts between nodes and the number of contents
generated over time. To do so, we calculated the correlation between the delta
in terms of similarity at each step of the simulations with respect to the
previous step and the number contacts and the number of contents at each
step. The results of this correlation analysis are reported in
Table~\ref{table:corr-values} in the third and fourth columns. In particular,
the two columns report the correlation values between the similarity delta ($Y$)
and the number of generated contents ($X1$) and the number of contacts between
agents ($X2$) respectively. The second column of the table reports the value for
the coefficient of determination ($R^2$) of the linear regression analysis
between the three measures, following the equation $Y=\beta_1 X_1 + \beta_2
X_2$. It is worth noting that, for all scenarios, the correlation and the $R^2$
values are rather low. This indicates that a simple analysis of the time series
of the number of contacts and contents over time, alone, is not sufficient to
describe the results of our simulations.

From Table~\ref{table:corr-values}, we note that the impact of the number of
contents generated is negatively correlated with the similarity of local and
global graphs for the WFD@Expo2015 scenario. This is perhaps not too surprising,
as a higher number of contents requires more contacts to disseminate the
generated knowledge. Interestingly enough, the effect is the opposite for KDD
and Helsinki, where there is a positive correlation between the similarity delta
and the number of contents. This might be a combined effect of the increasing
number of contacts in the hours of the day when people have a higher activity on
Twitter, as they probably coincide with the end of the working
activity. Nevertheless, the linear regression based on the combination of the
number of contact and contents is able to explain the variations in terms of
similarity. This confirms once again that the dynamics of the considered
scenarios cannot be simply described from aggregated measures, and the
simulations were necessary for the complete evaluation of p-PLIERS.
  
It is worth noting that p-PLIERS performs well even when knowledge is limited to
a very short time window in the past, at least for scenarios where a relatively
small geographic area is considered (WFD@Expo2015 and KDD scenarios). In fact,
nodes do not need much time for understanding what is happening around them in
these scenarios, and they can efficiently take decisions about currently
available contents. This is particularly relevant for opportunistic networking
scenarios, where nodes have very limited resources. For the urban scenario of
Helsinki, in which the considered area is much larger than that of the other
scenarios and the density of nodes is lower, the results indicate that a larger
time window is required for a good approximation of the global knowledge about
contents in the network. With a view to smart cities, a possible solution to
improve the diffusion of knowledge and the accuracy of p-PLIERS in this type of
scenario could be to exploit the public transportation system's nodes (e.g.,
buses, trams, or taxis) as additional information carriers.

\section{Possible Applications}
\label{sec:possible_applications}

The algorithm proposed in this paper can represent a general framework for the
development of opportunistic networking applications in several domains. From a
high level perspective, we identified two separate cases where the algorithm
could be useful for mobile applications in opportunistic settings. The former
involves autonomous content dissemination by the nodes, whereas the latter
requires the manual interaction of the user to explicitly download items in
specific services.

\subsection{Automatic Download for Content Dissemination and Routing}

In this case, the algorithm could be used as part of an autonomous decision
system for improving content dissemination services or routing algorithms in
opportunistic network settings. Specifically, each node could use our algorithm
to automatically decide which items are interesting for it or for other nodes in
the network. Then, it can take decisions on the data to download or on the route
that the messages must follow based on the calculated recommendations to improve
opportunistic routing algorithms. Of course, these decisions could be improved
by using information about physical contacts between nodes, as previously
proposed by Lo Giusto et al.~\cite{lo2010folksonomy}, and also additional
context information not necessarily related to the folksonomies used in this
work.

\begin{table}[t]
  \centering \footnotesize
\caption{Relation between graph similarity ($Y$) and (i) number of items
  generated at each simulation step ($X_1$), (ii) number of contacts at each
  step ($X_2$).}
\label{table:corr-values}
\begin{tabular}{l|r|r|r}
	\toprule Scenario & $R^2$ for fitting & $r_{YX_1}$ & $r_{YX_2}$ \\ &
        $Y=\beta_1 X_1 + \beta_2 X_2$ & & \\ \midrule Expo - 250 agents & 0.011
        & -0.017 & 0.117\\ Expo - 500 agents & 0.075 & -0.187 & 0.049\\ Expo -
        900 agents & 0.012 & -0.074 & 0.057\\ \hline KDD & 0.011 & 0.057 &
        -0.089\\ \hline Helsinki - 1 day & 0.032 & 0.175 & 0.072\\ Helsinki - 2
        days & 0.036 & 0.183 & 0.070\\ Helsinki - 3 days & 0.050 & 0.218 &
        0.047\\ \bottomrule
\end{tabular}
\end{table}

Note that the score calculated by recommender systems gives only a relative
weight to each item, but does not provide an absolute importance to them, apart
from excluding those items that receive a weight equal to 0. For this reason, in
order to evaluate new items encountered in the network, it could be useful for
the nodes to maintain a history of the scores received by other items seen in
the past, or to compare the scores of new items with those of already downloaded
items. An application based on this mechanism could evaluate the average
importance of the items seen in the past, considering a fixed time window, and
it can decide to download, for example, only the items that exceed this
average. Other possible variations of this scheme may be considered, of
course. For example, the node could download only the items exceeding a
percentile of the distribution of the scores of the items seen in the past
window. Alternatively, an application based on our algorithm may decide to
download the items as soon as it finds them from its neighbors, without
requiring to ``scan'' the network for a certain time before being able to decide
which items are interesting and which ones are not. This could nonetheless
require a buffer of items with limited size, which is possibly updated each time
a new neighbor is encountered. The buffer is initially filled in with all the
encountered items, but when the maximum size is reached, items are replaced by
new items with higher recommendation scores.

\subsection{Manual Download for File Sharing and Recommendation Services}

Other possible application scenarios may require the users to directly interact
with the algorithm to decide whether they want to download the recommended items
or not. For example, the algorithm could be used by applications that search for
multimedia files (e.g., songs or videos) from other peers, and use item
recommendations to decide which of these files may be interesting for the
users. The data download, in this case, could be performed by the user
directly. The history of downloaded items may be used to update the interests of
the users in the folksonomy graphs. More specifically, each time a user manually
downloads an item, we are sure that the item is interesting for her, and we can
thus add a link between the node representing the user in the folksonomy graph
and the node representing the downloaded item, allowing more personalized
recommendations in the future, based not only on the list of created items, but
also on the history of downloaded items.

\section{Conclusion}
\label{sec:conclusion}

In this work, we presented p-PLIERS, a novel distributed algorithm implementing
the PLIERS tag-based recommender system, which selects highly personalized
contents of interest for mobile users in opportunistic networking scenarios. The
algorithm is able to adapt to heterogeneous interest profiles of different
users, and effectively operates also when limited knowledge about the system is
maintained. It performs more accurate recommendations than other solutions
proposed in literature in terms of personalization with respect to the interests
of various users.

We validated the applicability of our proposal in real pervasive environments,
by simulating the use of PLIERS for content dissemination in three realistic
scenarios, a big event (WFD@Expo2015), a large conference (ACM KDD'15), and a
working day in a city center (Helsinki). In these scenarios, contents are
dynamically generated following the tweets created during the simulated time,
and each node knows only part of the whole dataset, namely, the local
information and the knowledge gathered from other nodes encountered while
moving. Furthermore, nodes have limited memory from which old knowledge is
purged. Also in this case, p-PLIERS proves to be able to provide effective
recommendations, comparable to those achievable if global knowledge were
accessible to nodes.

Future work consists in investigating other mechanisms to limit the knowledge
held and exchanged by nodes~--~while preserving recommendations accuracy~--~such
as the use of learning policies allowing nodes to discern and just maintain the
most significant information to compute appropriate suggestions. In addition, we
are currently working on the implementation of a prototypical application for
content dissemination in opportunistic networks, which will allow us to evaluate
the algorithm proposed in this paper in a real scenario.

\section{Acknowledgement}
This work has been partially funded by the EC under EIT Digital GameBus project
(Business Plan 2015).

\bibliographystyle{plain} \bibliography{paper}

\end{document}